\documentclass[12pt,preprint]{emulateapj}

\usepackage{lscape}

\begin{document}

\title{Multiple Star Formation to the Bottom of the IMF}

\author{Adam L. Kraus\altaffilmark{1,2} and Lynne A. Hillenbrand\altaffilmark{2}}

\altaffiltext{1}{Hubble Fellow; Institute for Astronomy, University of
Hawaii, 2680 Woodlawn Dr., Honolulu, HI 96822, USA}
\altaffiltext{2}{California Institute of Technology, Department of
Astrophysics, MC 249-17, Pasadena, CA 91125, USA}

\begin{abstract}

The frequency and properties of multiple star systems offer powerful tests of star formation models. Multiplicity surveys over the past decade have shown that binary properties vary strongly with mass, but the functional forms and the interplay between frequency and semimajor axis remain largely unconstrained. We present the results of a large-scale survey of multiplicity at the bottom of the IMF in several nearby young associations, encompassing 78 very low mass members observed with Keck laser guide star adaptive optics. Our survey confirms the overall trend observed in the field for lower-mass binary systems to be less frequent and more compact, including a null detection for any substellar binary systems with separations wider than $\sim$7 AU. Combined with a Bayesian re-analysis of existing surveys, our results demonstrate that the binary frequency and binary separations decline smoothly between masses of 0.5 $M_{\sun}$ and 0.02 $M_{\sun}$, though we can not distinguish the functional form of this decline due to a degeneracy between the total binary frequency and the mean binary separation. We also show that the mass ratio distribution becomes progressively more concentrated at $q\sim$1 for declining masses, though a small number of systems appear to have unusually wide separations and low mass ratios for their mass. Finally, we compare our results to synthetic binary populations generated by smoothed particle hydrodynamic simulations, noting the similarities and discussing possible explanations for the differences.

\end{abstract}


\section{Introduction}

The frequency and properties of multiple star systems offer powerful constraints on star formation and early cluster evolution. The semimajor axis of a binary system should correspond to the characteristic size of its progenitor core at the time of fragmentation, so the binary separation distribution constrains the range of sizes and the size evolution for cores\citep[e.g.,][and references therein]{Sterzik:2003fx}. The overall binary frequency and the mass ratio distribution are set by the detailed physics of binary fragmentation\citep[][]{Delgado-Donate:2004id}, and each binary system's mass ratio will depend on the post-fragmentation accretion history\citep[][]{Bate:1997kx}, while formation in environments with high stellar density could shape the binary population as and after it forms\citep[][]{Kroupa:1999fd}. A successful model for star formation should be able to match the observed frequency and properties of the binary star population, as well as any mass-dependent changes in these parameters.

The past two decades have seen numerous studies of nearby field binary systems in order to constrain their frequency and properties. These surveys \citep[e.g.,][]{Duquennoy:1991zh,Fischer:1992if,Close:2003ud,Bouy:2003qc,Burgasser:2003mw} have found that binary frequencies and properties are very strongly dependent on mass. Solar-mass stars have high binary frequencies ($\ga$60\%) and maximum separations of up to $\sim$10$^4$ AU. By contrast, M dwarfs have moderately high binary frequencies (30--40\%) and few binary companions with separations of more than $\sim$1000 AU, while brown dwarfs have low binary frequencies ($\sim$15\% for all companions with separations $\ga$2--4 AU) and few companions with separations $>$10 AU.

However, field multiplicity results face unavoidable ambiguity near and below the substellar regime. These surveys provide only weak constraints on the mass dependence of substellar binary properties due to the degeneracy between brown dwarf masses and ages, and mass ratios are similarly difficult to estimate. Also, the field represents a composite population drawn from all star-formation regions, so field surveys cannot probe the dependence of binary properties on initial conditions (the stellar density or total mass) and evolutionary history (the degree of dynamical evolution each system undergoes before leaving its natal environment). For example, the separation distribution for binary systems is truncated at separations of $\sim$100 AU in open clusters like Praesepe \citep[e.g.,][]{Patience:2002qh}, whereas unbound young associations have binary systems as wide as 10$^4$ AU \citep[][]{Kraus:2008fr,Kraus:2009uq}.

These complications have motivated a large number of multiplicity surveys in nearby star-forming regions and young clusters. Several survey programs have found that the solar-mass stars in loosely-bound young associations have extremely high binary frequencies\citep[][]{Ghez:1993xh,Leinert:1993gd,Simon:1995yi,Kohler:2000lo,Kraus:2008zr}. The binary frequency in young open clusters appears to be significantly lower\citep[e.g.,][]{Petr:1998ys,Kohler:2006wy}, which could be interpreted either as early dynamical evolution or a signature of the different primordial environment. Surveys of very low-mass stars and brown dwarfs have concentrated mainly on nearby unbound associations \citep[][]{Kraus:2005qf,Kraus:2006ul,Konopacky:2007pd,Ahmic:2007hn,Biller:2011mw}, but produced results that largely match the field: low-mass binary systems are rare and tend to have small separations and similar component masses.

The aforementioned surveys of low-mass multiplicity in young associations used very modest sample sizes since high-resolution imaging techniques were observationally expensive. As a result, their tentative conclusions raised as many questions as they answered. The handful of binary systems they discovered tended to fall in the upper end of the surveys' mass ranges, with very few binary companions to genuinely substellar primaries. This suggested that the binary frequency might decline with primary mass through this range, an observation that is difficult to test among low-mass field binaries.

The limited sample sizes and heterogeneous nature of previous surveys have prohibited any detailed analysis of the mass dependence of multiple star formation, especially in the low-mass regime ($M\la$0.15 $M_{\sun}$) where mass-dependent effects seem to be most significant. To address this shortcoming, we present a large-scale survey of multiplicity at the bottom of the IMF in several nearby young associations. In Section 2, we list our survey's sample and describe our survey's observations, and in Section 3, we explain the analysis techniques used in our program. In Section 4, we describe the results of our observations. Finally, in Section 5, we use our results and other results from the literature to constrain the mass-dependent properties of low-mass multiple star formation and compare those properties to the results of theoretical models.

\section{Sample and Observations}

\subsection{Sample Selection}

Nearby star-forming regions have been the target of numerous wide-field photometric imaging surveys to detect new low-mass members \citep[e.g.,][]{Luhman:2004yq,Luhman:2006dp,Slesnick:2006pi,Slesnick:2006xr,Slesnick:2008mi}. These surveys identified candidate members based on their location on an optical or near-infrared color-magnitude diagram, and membership was then confirmed spectroscopically via the detection of lithium absorption, excess $H\alpha$ emission, or low surface gravity, all of which are indicators of youth. We chose to concentrate on Taurus-Auriga \citep[$\tau \sim$1--2 Myr; $d \sim 145$ pc;][]{Kraus:2009fk,Torres:2009ct} and Upper Scorpius \citep[$\tau \sim$5 Myr; $d \sim 145$ pc;][]{de-Zeeuw:1999xv,Preibisch:2002qt}because they are the nearest young associations that are accessible from the northern hemisphere. In our analysis, we also use results from the literature on the Cha-I association \citep[$\tau \sim$2--3 Myr; $d \sim 185$ pc;][]{Luhman:2004yq}, which is similar to Taurus in age and environment. All regions have very low stellar densities ($N \la$10 pc$^{-3}$), which is critical for minimizing the complicating role of dynamical interactions.

Our initial observational sample included all late-type members of each association (SpT$\ge$M4) that had been identified by 2006 and that hadn't been observed at high angular resolution. However, as we describe in Section 3, we lost a significant fraction of our observing time to poor weather and instrument problems. This left our Taurus sample significantly incomplete for members discovered in 2006, plus we were unable to observe three members that had been identified earlier (J1-4423, V410 X-ray 6, and 2MASS J04163049). The effect on Upper Sco was to limit our observed sample to only the latest-type members (SpT$\ge$M6.5). We also were unable to observe 10 Taurus members that had no suitable tip-tilt stars available. This omission introduces a bias against the most reddened members of Taurus since the most heavily extincted stars were least likely to have an optically bright star nearby that could serve as a tip-tilt reference. The density of field stars is very high in Upper Sco, plus the association is almost completely cleared of its primordial molecular material, so we were always able to find a suitable tip-tilt star.

We have supplemented this observational sample with the results of numerous previous multiplicity surveys. Taurus has been a very popular target for multiplicity surveys, and association members in our spectral type range have been observed with speckle interferometry \citep[][]{Ghez:1993xh,Konopacky:2007pd}, lunar occultations \citep[][]{Simon:1995yi}, HST imaging \citep[][]{Padgett:1999rt,White:2001jf,Kraus:2006ul}, AO imaging \citep[][]{Correia:2006pi}, and aperture-masking interferometry \citep[][]{Kraus:2011tg}. Upper Scorpius has been the subject of several surveys as well, and members have been observed with speckle interferometry \citep[][]{Kohler:2000lo}, HST imaging \citep[][]{Kraus:2005qf}, and AO imaging and aperture-masking interferometry (Kraus et al. 2008). In most cases (and almost certainly in aggregate), the sample members were selected seemingly at random. As a result, we adopt the combined set as an unbiased sample.

Finally, we also observed a small number of other targets that fall outside these selection parameters, but were considered interesting for other reasons. In both associations, we observed a number of candidate wide binary systems that seemed to have unusually low binding energies. We already described the observations for UScoJ1606-1935 in a previous paper \citep[][]{Kraus:2007gf}, and we reported the astrometric measurements for the rest in our paper on wide binary formation \citep[][]{Kraus:2009uq}. In this paper, we report on the search for higher-order multiplicity. We also observed several candidate Taurus members discovered by \citep[][]{Slesnick:2006xr} that are not part of the young Taurus population, but might represent an older, more widely distributed population of young stars and brown dwarfs. Finally, we observed the known binary V928 Tau because it served as the tip-tilt reference for CFHT-Tau-7 and we typically imaged tip-tilt references for a data quality check. However, we do not report any results for CFHT-Tau-7 because the observing conditions at the time were too marginal for adaptive optics to yield any meaningful correction.

In Table 1, we list the young association members that we observed in our study. The $K$ magnitude for each target was taken from 2MASS (Skrutskie et al. 2006), while the $R$ magnitude and distance to each star's tip-tilt reference are from the USNO-B1.0 catalog (Monet et al. 2003). We also list references for the handful of objects which have also been observed in other high-resolution imaging surveys; in several cases, our detection limits for small separations were superceded by the survey by \citet[][]{Konopacky:2007pd}, so we adopted those limits where appropriate.

\subsection{Observations}

Most of the data that we summarize were obtained in 4 observing runs, totaling 10 nights, between December 2005 and March 2007. One source was observed during a time trade in December 2006. Most of our observations were obtained using laser guide star adaptive optics \citep[LGSAO;][]{Wizinowich:2006zn} on the Keck-II telescope with NIRC2 (K. Matthews, in prep), a high spatial resolution near-infrared camera. During some periods of moderate cloud cover that were not suitable for laser operation, we also used natural guide star adaptive optics (NGSAO) to observe sample members with very close and bright tip-tilt stars. In the worst conditions, we also observed some higher-mass stars that did not fall in our sample; most of these observations have been described in our previous papers \citep[][]{Kraus:2008fr,Kraus:2009uq}, so we report the rest here for completeness.

The weather conditions were highly variable over the course of our campaign, with only five nights suitable for laser operations. Several of the remaining nights were also impacted by poor seeing. Despite numerous difficulties, we report 82 observations of young association members with LGSAO and 5 additional observations with NGSAO, encompassing 78 different targets. In Table 2, we summarize the observations for each target. We also list the typical PSF FWHM for each target; a significant fraction of the targets used bright on-axis tip-tilt stars that should have achieved diffraction-limited performance (strehl$\sim$30\%), but we only achieved this performance on two nights.

All of the images presented here were produced with the narrow camera, which has a field of view of 10.2\arcsec and a pixel size of 9.963 mas pix$^{-1}$ \citep[][]{Ghez:2008my}.  All targets were observed with the $K_p$ filter; in most cases, we did not obtain observations in other filters because most background stars have $J-K$ and $H-K$ colors that are not sufficiently different from young stars as to allow secure identification. Many results we draw from the literature appear to have been observed with $K$ or $K_s$ filters (though some do not specify), but we treat all three filters equivalently since the color terms are smaller than the typical photometric uncertainties (e.g., Carpenter et al. 2002). During early observing runs, we used a three-point dither pattern that was designed to avoid the bottom-left quadrant, which suffers from high read noise. After February 2006, we obtained all of our observations in a diagonal two-point dither pattern because experience showed that dithers degrade the AO correction until several exposures have been taken with the Low-Bandwidth Wavefront Sensor. The delay before returning to optimal correction represented a significant overhead that we sought to minimize.

Many of the targets are relatively bright in the NIR and require very short integration times to avoid saturation or nonlinearity, so a large fraction of our observations were taken in correlated double-sampling (CDS) mode, for which the array read noise is 38 electrons read$^{-1}$. Where possible, we observed targets in multiple correlated double-sampling (MCDS) mode, where multiple reads are taken at the beginning and ending of each exposure; this choice reduces the read noise by approximately the square root of the number of reads. This is doubly significant because the read noise per coadd and the total number of coadds per exposure are both reduced. In all cases, the read noise is negligible compared to PSF variations from the primary at separations of $<$1\arcsec. However, it always dominated over the sky background in determining our faint-source detection limits at large separations from the science target. In all cases, the images were flat-fielded and dark- and bias-subtracted using standard IRAF procedures.

\section{Analysis Methods}

\subsection{Source Identification and Detection Limits}

Source identification in AO imagery is a complicated endeavour. In NGSAO mode, the gross shape of the PSF depends on the target's optical brightness and the seeing, while the fine structure is determined by speckle patterns that continuously change on timescales ranging from seconds to hours. The LGSAO PSF is further complicated by variations in laser return strength, tip-tilt anisoplanatism with respect to off-axis guide stars, and heightened sensitivity to telescope effects like wind shake. Finally, observations in poor weather are further complicated by rapid PSF quality variation due to changing atmospheric conditions. The source detection process can be divided into two regimes: a wide regime where the PSF core is negligible and speckle confusion dominates (projected separation $\rho$$\ga$2 times the core FWHM), and a close regime where shape and width variations in the PSF core dominate and speckle confusion is negligible. We have adopted a different method in each separation regime.

\subsubsection{The Wide, Speckle-Dominated Regime}

As was summarized by \citet[][]{Metchev:2009hh}, there are four common methods used to subtract the primary star's flux and identify companions in AO imagery: subtracting a median PSF representing all similar observations, subtracting a 180$^o$ rotated version of the same image, high-pass filtering by subtracting a Gaussian-smoothed version of the same image, or subtracting the azimuthally-averaged profile. We conducted experiments with these techniques, but we found that the speckle mitigation strategies that are vital for high-strehl NGSAO data are actually only marginally useful for low-strehl LGSAO data. Most of the flux that would be found in discrete speckles in high-strehl data is instead averaged into the seeing-limited halo, rendering the PSF less azimuthally variable at a given separation. The brightest speckles remain distinguishable, but at far lower contrast with respect to the surrounding median flux. Since the noise floor is brighter and the noise ceiling is fainter, there is less to be gained from exceeding the noise ceiling.

The sparsity and relatively low contrast of speckles in LGSAO data suggest that a different strategy is optimal for our data. Speckle mitigation and subtraction of the primary star's flux are observationally expensive, so the preferred strategy should be to characterize the mean and standard deviation of the brightness distribution of the PSF as a function of separation, then set the source detection limit above the expected ceiling for speckle brightness. We characterized the brightness distribution of each target's PSF by measuring the flux through photometric apertures placed at a range of separations and PAs, then measuring the mean and standard deviation for all apertures in a given bin of separation. The apertures were placed on a rectangular grid with spacing of 25 mas in order to ensure that the small number of speckles were detected, and the aperture sizes were matched to the FWHM of the PSF core for the primary. We measured this aperture photometry using the IRAF\footnote{IRAF is distributed by the National Optical Astronomy Observatories, which are operated by the Association of Universities for Research in Astronomy, Inc., under cooperative agreement with the National Science Foundation.} task PHOT, which is distributed as part of the DAOPHOT package \citep[][]{Stetson:1987ya}.

In Figure 1, we show the contrast as a function of separation for three stars that span our survey's data quality, as well as the 5$\sigma$ envelope for each source. We found that there was typically 1 detection at 4--5$\sigma$ per 2 stars, one detection at 5--6$\sigma$ per 10 stars, and no detections among any of our targets at 6--10$\sigma$. This indicates that a 6$\sigma$ clip should be uncontaminated by spurious detections, while a 5$\sigma$ clip can be adopted if the few remaining speckles can be confidently identified as such. We found that all of the 5--6$\sigma$ candidate detections fell on the PSF's diffraction spikes, were sufficiently short-lived so as to not appear in all observations of a target, or were sufficiently long-lived so as to appear in observations of multiple sequential targets. We therefore suggest that all such 5--6$\sigma$ candidate detections are spurious, and adopt a 5$\sigma$ clip as our survey's detection limits. All candidate detections that sit well above the 5$\sigma$ limit appear to be genuine astronomical sources, though not necessarily comoving companions; we will revisit this distinction in Section 5.2.

 \begin{figure}
 \plotone{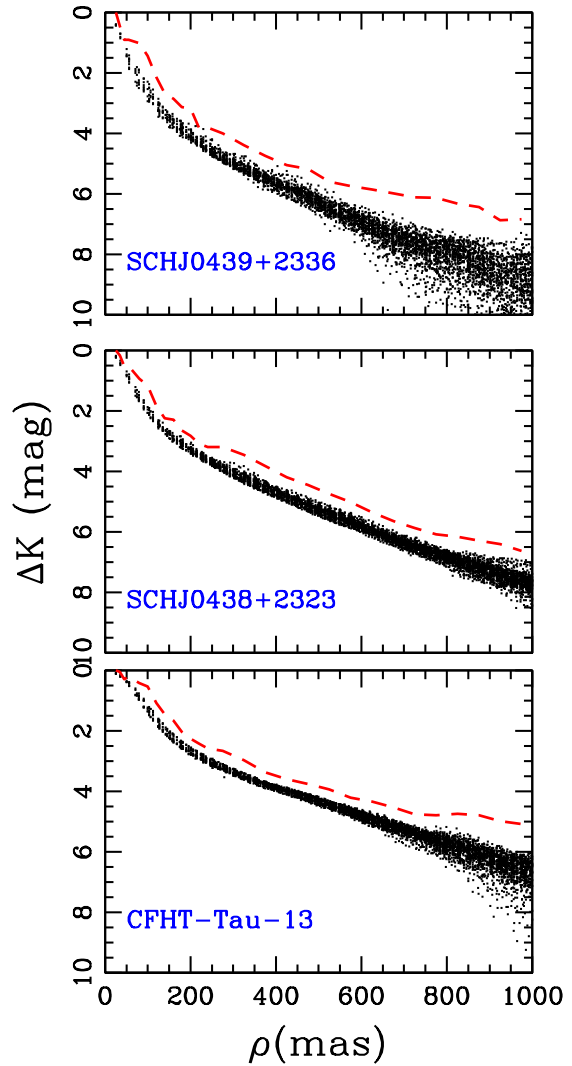}
 \caption{Contrast limits at wide separations ($>$100 mas) for three representative stars in our sample: SCH J0439016+2336030 (top), SCH J0438587+2323596 (middle), and CFHT-Tau-13 (bottom). The small black points show the flux as a function of separation for apertures placed at a range of separations and PAs from the primary, while the red dashed line shows the +5 sigma envelope above these points. A handful of candidate sources fall above this significance level, but all can be identified as speckles, so we have defined this envelope for each target and use it as our survey's detection limit. The main difference in contrast is determined by the quality of the tip-tilt reference star (Table 1); tip-tilt references which are brighter or located on-axis yield better AO correction.}
 \end{figure}

\subsubsection{The Close, Core-Dominated Regime}

For separations near the PSF FWHM, the detection limits are driven by time- and spatially-dependent variations of the shape of the PSF core. LGSAO observations seem to be more susceptible to all of the weather effects that can degrade NGSAO observations, so distinguishing genuine companions from PSF artifacts is a significant challenge. The primary effect we see is for wind shake to cause PSF elongation in the direction of the zenith, perhaps because the tip-tilt sensor and low-bandwidth wavefront sensor operate at a lower rate and can't fully sample high-frequency oscillations of the telescope. Tip-tilt anisoplanatism is also significant for observations with off-axis tip-tilt guide stars, an effect that becomes worse in poor seeing because the isokinetic angle becomes smaller. This causes elongation along the position angle to the tip-tilt star. Finally, significant variations in the AO correction cause the PSF FWHM itself to vary by a factor of $\sim$3 across our sample; a few of the lowest-quality observations have a PSF FWHM approaching 150 mas.

We have characterized these effects by fitting each science target's PSF core with a bivariate Gaussian distribution. This fit directly yields the PSF elongation (the ratio of the major axis $a$ and minor axis $b$) and its direction (the PA of the major axis). For each our targets, we report the minor axis FWHM (i.e. the resolution prior to elongation effects) and the fractional elongation in Table 2. If windshake and tip-tilt anisoplanatism are the dominant sources of PSF asymmetry, then most sources should have a PA that is preferentially aligned with either the tip-tilt angle or the zenith angle. A set of selection criteria based on these quantities also lends itself to rigorous characterization of the detection limits, as artificial star tests can be used to determine whether a system with given separation and contrast (and perhaps PA) would be detected.

In Figure 2 (top), we plot the PSF elongation and the relative angle between the PSF and the zenith angle ($|\theta$$_{PSF}$-$\theta$$_{zen}|$) for all science targets which served as their own tip-tilt references. Of the 21 sources which are not independently-confirmed binary systems, 9 are aligned to within $<$10$^o$ of the zenith angle and 7 of the remaining 12 are aligned to within 10--45$^o$. This strong trend indicates that wind-induced elongation was significant across the majority of our sample, even on those nights with moderate winds. Also, all of the targets that are not independently-confirmed binary systems have elongations of $<$30\%, which seems to be the ceiling for PSF elongation due to telescope or atmospheric effects.

In Figure 2 (bottom), we consider the rest of our sample in plotting the PSF elongation versus the minimum of the relative angle either between PSF and zenith ($|\theta$$_{PSF}$-$\theta$$_{zen}|$) or between PSF and tip-tilt ($|\theta$$_{PSF}$-$\theta$$_{TT}|$). These targets also show a pronounced tendency to align with either the zenith or tip-tilt, though the result is more complicated because many targets have a net elongation intermediate between the two directions. There are fewer confirmed binary systems for comparison, but most of the targets fall below elongations of 40\%, suggesting that this is the ceiling for combined wind and tip-tilt effects. Many of the targets with elongation $>$40\% fall among our poorest sample and have significantly different elongation angles and magnitudes in each exposure, while all of the remainder present consistent and apparently double-peaked PSFs.

In light of these trends, we have adopted two criteria for identifying a source as a candidate binary system. First, it must have a PSF elongation of $>$40\%, which appears to be the ceiling for any weather-based effect in all but the worst data. Second, similar PSF elongation must not be seen for other sources in the science target FOV, in our preliminary image of the tip-tilt guide star, or in observations of the previous or subsequent science target. Finally, the astrometry and photometry for a fit of 2 point sources must be consistent across all exposures; the poor data with elongations $>$40\% tends to vary its PSF shape on extremely short timescales, yielding extremely inconsistent fits across the full dataset. We have inferred the source detection limit for each of our targets by measuring its minor-axis FWHM, then using artificial star tests to determine what ranges of companion separation and companion bright would have elongated a circular PSF with that FWHM to $>$40\%.

Finally, there are also some cases where companions can be confidently studied below our survey's detection limit, such as if the companions were previously identified in another survey (i.e. V410 X-ray 3) or if a third bright star can be used as an independent PSF calibrator (UScoJ1607-2019). We have used PSF-fitting techniques (Section 3.2.3) to recover the photometry and astrometry for these close binary pairs, though we generally cannot include them in our statistics if their detection relied on a special feature of the system like high-order multiplicity. We also note that we were unable to recover accurate astrometry and photometry for MHO-Tau-8, which suggests that its orbital motion might have carried it inward from its last-known projected separation \citep[$\sim$40 mas;][]{Kraus:2006ul}.

\begin{figure}
 \epsscale{0.9}
 \plotone{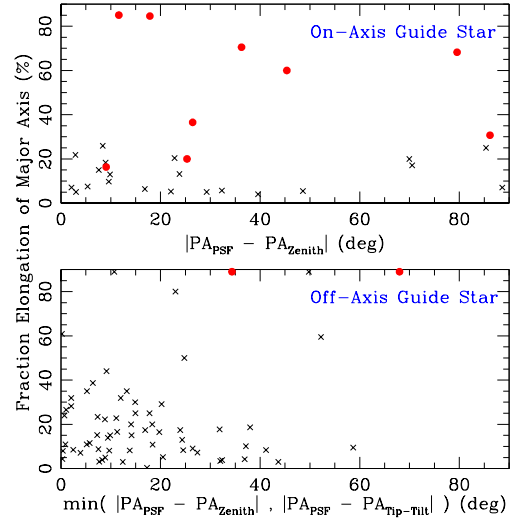}
 \caption{Top: Fractional PSF elongation as a function of its alignment with the direction to zenith for independently-confirmed binary systems (red circles) and all other targets (black crosses) that served as their own tip-tilt guide star. Most non-binary sources appear to be roughly aligned with zenith, a trend which indicates that wind-induced telescope shake is common among our observations. All sources which are not known binary systems appear to have PSF elongations of $<$30\%, which seems to be the ceiling for weather-induced effects. Bottom: A similar figure for targets which had off-axis tip-tilt guide stars, where we plot the elongation of the target PSF with respect to the closer of the angle to zenith or the angle to the tip-tilt. Tip-tilt anisoplanatism also seems to induce PSF elongation, but with the exception of firmly-detected binary systems and targets with very poor data quality, the ceiling for observational effects is $<$40\%.}
 \end{figure}

\subsection{Photometry and Astrometry}

We measured relative photometry and astrometry for candidate companions using the IRAF package DAOPHOT \citep[][]{Stetson:1987ya}. For source pairs with small separations, where the two PSF cores were not unambiguously resolved, we used the PSF-fitting ALLSTAR routine. For pairs with wider separations, we used the aperture photometry package PHOT. We analyzed each frame separately in order to estimate the uncertainty from the scatter between all frames; this also allowed us to reject some frames with subpar AO correction. Our final results represent the mean and standard deviation for all observations.

For the close binaries that we analyzed with ALLSTAR, we reconstructed the single-star PSF out of the merged binary PSF using the algorithm described in \citet[][]{Kraus:2007gf}, which iteratively fits a template PSF to the primary and then subtracts the secondary to fit an improved estimate of the primary. For one triple system, UScoJ1607-2019, we chose instead to use the single secondary as a PSF template for fitting the close pair constituting the primary. This choice allowed us to clearly distinguish the close pair despite a separation ($\sim$50 mas) that was significantly lower than the PSF FWHM ($\sim$70 mas).

Our relative astrometric measurements were distortion-corrected using a new high-order distortion solution (Cameron 2008) that delivers a significant performance improvement as compared to the solution presented in the NIRC2 pre-ship manual\footnote{\url{http://www2.keck.hawaii.edu/realpublic/inst/nirc2/}}. This distortion solution was derived from observations of a pinhole mask in the NIRC2 filter wheel, so it does not include any distortions introduced upstream of this point. The remaining residuals due to these uncorrected distortions are $\sim$5 mas for positions separated by $\sim$5--10\arcsec\, (J. Lu, priv. comm.). We calibrated our photometry using the known 2MASS $K_s$ magnitudes for each of our science targets; these absolute magnitudes are uncertain by $\sim$0.1--0.2 magnitudes due to the intrinsic variability of young stars (resulting from accretion or rotation).

\subsection{(Sub)stellar and Companion Properties}

Stellar properties can be difficult to estimate, particularly for young stars, since pre-main-sequence stellar evolutionary models are not well-calibrated. The model-predicted masses of young stars could be systematically uncertain by as much as 20\% \citep[e.g.,][]{Hillenbrand:2004bh}, and estimates for individual stars could be uncertain by factors of 2 or more if their observed luminosities are biased by unresolved multiplicity or the intrinsic variability that young stars often display (from accretion or from rotational modulation of star spots). This suggests that any prescription for determining stellar properties should be treated with caution.

We estimated the properties of all of our sample members using the methods described in \citet[][]{Kraus:2007ve}. This procedure combines the 2 or 5 Myr isochrones of \citet[][]{Baraffe:1998yo} and \citet[][]{Chabrier:2000sh} with the temperature scales of \citet[][]{Schmidt-Kaler:1992ab} and \citep[][]{Luhman:2003pb} to directly convert observed spectral types to masses. Relative properties (mass ratios $q$) for all binaries in our sample were calculated by combining these isochrones and temperature scales with the empirical NIR colors and K-band bolometric corrections of \citet[][]{Kraus:2007mz} to estimate $q$ from the observed flux ratio $\Delta K'$.

For all binary systems, we have adopted the previously-measured(unresolved) spectral type for the brightest component and inferred its properties from that spectral type. This should be a robust assumption since equal-flux binary components will have similar spectral types and significantly fainter components would not have contributed significant flux to the original discovery spectrum. Projected spatial separations are calculated assuming the mean distance to the associations \citep[$\sim$145 pc for both Upper Sco and Taurus;][]{de-Zeeuw:1999xv,Torres:2009ct}. If the total radial depth of each association is equal to its angular extent ($\pm$8$^o$ or $\pm$20 pc), then the unknown depth of each system within the association implies an uncertainty in the projected spatial separation of $\pm$14\%. The systematic uncertainty due to the uncertainty in the mean distance of each association is negligible in comparison ($\la$2\%).

\section{Results}

\subsection{Candidate Companions and Detection Limits}

Our search for sources in the speckle-dominated regime (at separations $\ga$1.5 times the PSF FWHM and extending to the edge of the detector; Section 3.1.1) yielded 45 candidate companions among the 78 young stars and brown dwarfs in our observed sample. All candidates within $\la$1\arcsec of the target sit well above the 5$\sigma$ detection limit, so they all represent secure detections and do not seem to be spurious structures in the primary star's PSF. We also found numerous possible detections with significance levels of 5--6$\sigma$ in this regime, but as we described in Section 3.2.1, all of these possible detections appear to be spurious. Our corresponding search for sources in the core-dominated regime (at separations of order the PSF FWHM; Section 3.1.2) yielded 9 targets with PSF cores consistently elongated by $>$40\%. Many targets had PSF elongations below this limit, but as we discussed above, most appear to be distorted due to observational effects and not the presence of a companion. We address the membership probabilities of these candidates in Section 4.2.

In Table 3, we list our survey's candidate companions and report their flux ratios, separations, and position angles. We also plot the flux ratio $\Delta$$K'$ and the candidate companion brightness $K'$ as a function of separation in Figures 3 and 4, respectively. Finally, in Figure 5, we show contour plots for the 12 candidate companions with separations $<$1\arcsec\, and flux ratios of $\Delta$$K'<4$; we do not show contour plots for wider systems because they are completely resolved, and we do not show any faint companions because few are likely to be bound companions. The vast majority of our wide-separation candidates are near the detection limits of our survey, in the brightness range where a real companion would fall below the deuterium-burning limit ($\sim$13 $M_{Jup}$), so we expect that almost all are unassociated field stars. However, a handful of planetary-mass companions have been identified around young stars and brown dwarfs \citep[e.g., 2M1207b;][]{Chauvin:2004bd}, so we must consider the possibility that some of the candidates in our survey are also extremely low-mass companions. We will present second-epoch observations that test the membership of these candidates in a future paper. In all cases, the high astrometric precision of Keck/NIRC2 ($\sigma \sim$1--2 mas) and the relatively large proper motions ($\mu \sim $20--30 mas/yr) allow for a test of common proper motion using observations from consecutive observing seasons.

Of the 54 combined sources from the speckle-dominated and core-dominated regimes, 11 had already been identified as candidate binary companions by past survey efforts, so we note these past identifications in Table 3. Three of our candidate companions merit special attention. V410 X-ray3 AB was identified as a candidate binary in our previous HST/ACS multiplicity survey based on a marginal elongation of its PSF, but the best-fit separation was well inside the HST diffraction limit for the $i'$ and $z'$ filters. By contrast, the system is almost resolved in our $K'$ observations, and corresponding $JH$ images (Kraus 2009) clearly reveal V410 X-ray3 to be a genuine binary system. Two wide binary systems, USco 80 AB and UScoJ1607-2019 AB, were resolved to be hierarchical triples. USco 80 A is clearly resolved to be a close pair, while UScoJ1607-2019 A is marginally resolved.

The other 26 members of our sample have no resolved neighbors within our detection limits. These detection limits, which we derived using the methods described in Section 3.1 and list in Table 4, are extremely heterogeneous due to the wide range in observing conditions during our observing campaign. Some observations nearly achieve the expected limits for diffraction-limited images, while images from most of the other nights achieved significantly poorer conditions. We also note that our nominal detection limits for bound companions at wide separations was constrained by our followup efforts. We found many more faint candidates than we were able to follow up with second-epoch imaging, so we can not claim completeness beyond the maximum separation or flux ratio at which we have identified all of the field stars by testing for common proper motion.

 \begin{figure}
 \epsscale{1.0}
 \plotone{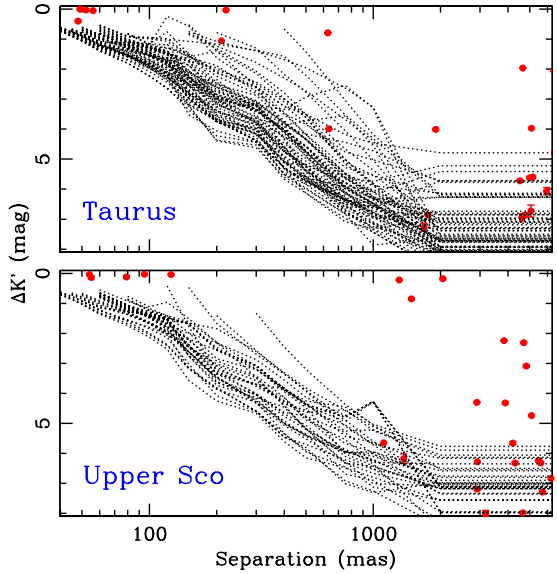}
 \caption{Separation and flux ratio for each of the candidate companions in our sample. The top panel shows our results for the 53 Taurus members in our sample, while the bottom panel shows our results for the 28 Upper Sco members. Red circles denote the candidate companions that we have detected, while dotted lines show the inferred detection limits for all sources.}
 \end{figure}

 \begin{figure}
 \epsscale{1.0}
 \plotone{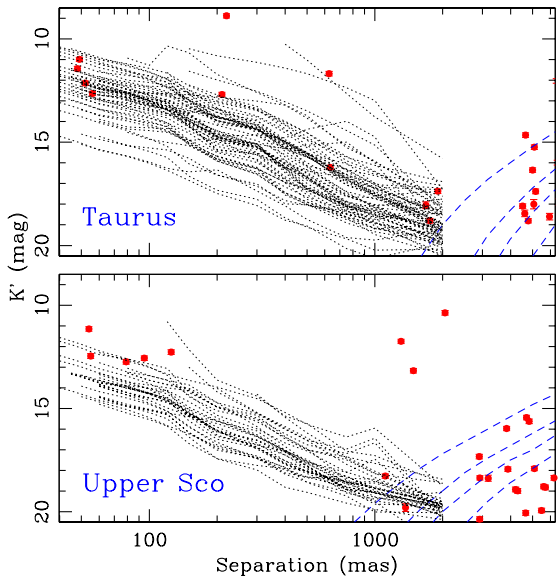}
 \caption{Separation and apparent magnitude $m_{K'}$ for each of the candidate companions in our sample. Red circles and dotted lines are defined as in Figure 3; blue dashed lines denote levels of constant contaminant density where we expect to find 1, 3, 5, or 10 background stars that are brighter and located at smaller projected separation.. We inferred these contamination rates using the star count models that we describe in Appendix A. All detection limits converge to the read-noise limit at separations of $>$2\arcsec, so we do not extend the limits beyond that separation since the lines would obscure the faint sources and the background contamination contours.}
 \end{figure}

 \begin{figure*}
 \epsscale{1.0}
 \plotone{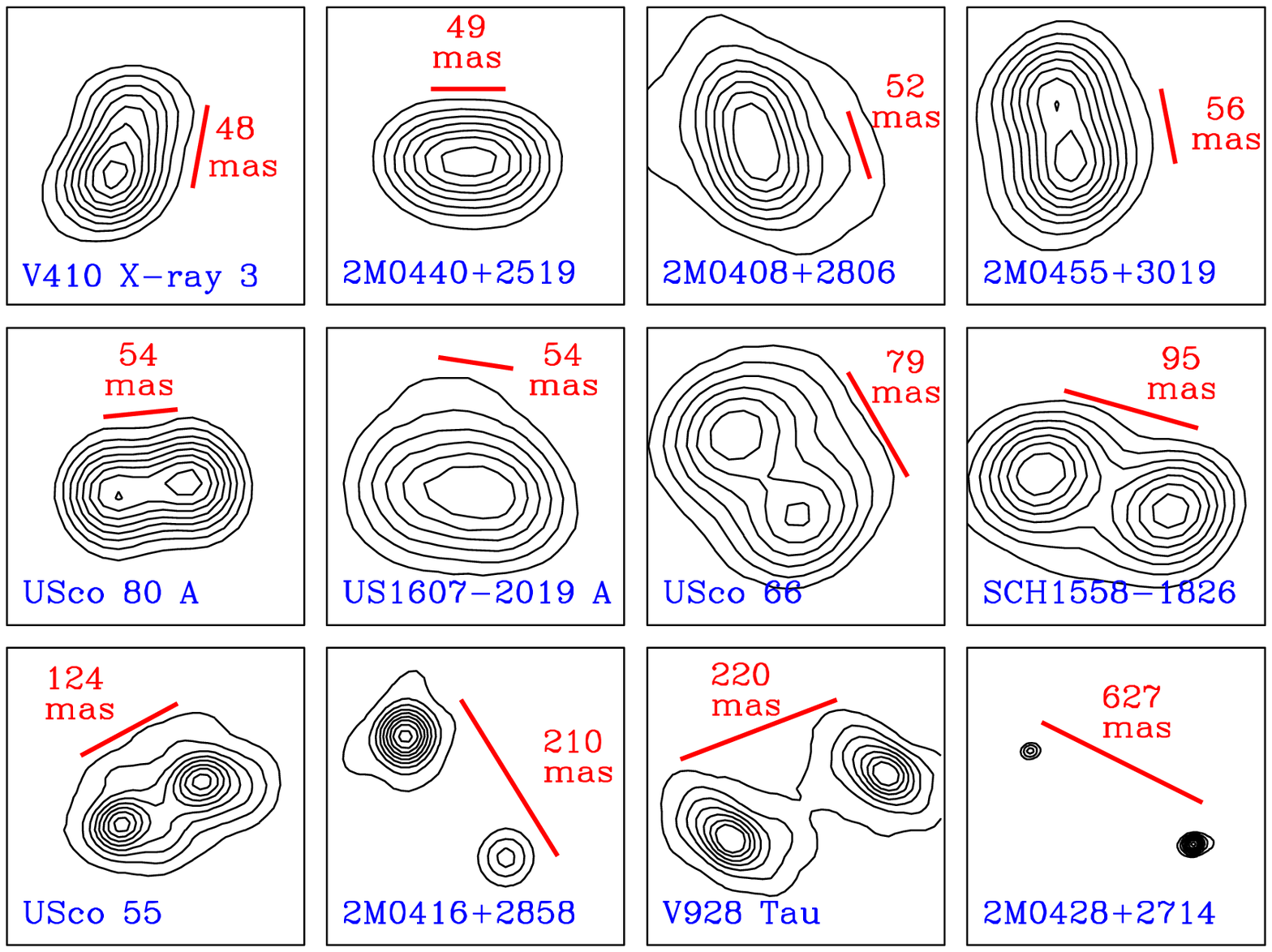}
 \caption{Contour plots for the twelve binary systems we observed to have separations of $<$1\arcsec. The top row includes four close binaries ($<$100 mas) in Taurus, the middle row includes four close binaries in Upper Sco, and the bottom row includes four wider binaries ($>$100 mas) from both associations. In each case, we plot contours at 95\% at the peak flux, and then in 10\% increments until reaching the seeing-limited halo; images with poor AO correction therefore show fewer contours}
 \end{figure*}

\subsection{Binary Systems and Field Stars}

Companion searches must address the prospect of chance alignments with background stars, especially surveys with very deep detection limits. We are obtaining multi-epoch astrometry for several candidates in order to test for common proper motion, but we can use statistical arguments to determine which candidate companions require those followup observations. As we describe more fully in Appendix A, we have updated the Milky Way model of \citet[][]{Bahcall:1980lr} to predict star counts as a function of magnitude for the line of sight toward each of our targets; these models allow us to predict the field star contamination rate and thereby determine which companions have a significant probability of being background stars.

In Figure 4, we plot the joint magnitude-separation limits at which our models predict we should find 1, 3, 5, or 10 background stars among all the targets observed in that association. In both associations, our models predict that we should find $<$1 background star with $K\la$15 within $<$5\arcsec, which suggests that all of the bright sources we observe well inside this limit are genuine companions. This limit agrees with our estimate based on 2MASS source counts in the direction of Taurus and Upper Sco \citep[][]{Kraus:2007ve}, which found that all neighbors down to the 2MASS 10$\sigma$ limit ($K=14.3$) could be assumed to be bound association members out to separations of 3--5\arcsec.

The status of our fainter candidates is not as clear. Our models predict that we should find only one background star with separation $\rho$$<$2\arcsec\, and brightness $K<$19 in Taurus, so the candidates inside this limit all seem very promising as potential analogs to 2M1207b. The background star density is higher in the direction of Upper Sco since it lies in the direction of the galactic center, so even with our smaller sample size, we still expect 3 chance alignments with $\rho$$<$2\arcsec and $K<$19. However, star count models are not well-constrained at $K\ga$14 since the counts are dominated by low-mass halo stars, a population that is not as well-studied as the brighter disk stars. As a result, the contamination rate for faint stars could be uncertain by a factor of at least 2--3. It would be prudent to measure common proper motion for any candidate companion fainter than $K\sim$14, especially since any genuine companion would have $M<$10--20 $M_{Jup}$, making it an extremely compelling discovery. The predicted background star contamination rate rises quickly for separations $\ga$3\arcsec, matching the many candidates we have discovered, so we provisionally adopt this separation as an outer limit at which it is worthwhile to test candidates for possible association. Future searches at wider separations should use seeing-limited data from publicly available surveys like UKIDSS, which will observe the majority of these targets in four NIR filters.

Finally, multiplicity surveys in young clusters and associations must also consider chance alignments between two unbound member stars. These chance alignments are extremely difficult to distinguish from genuine binary systems since all association members are young, at similar distance, and comoving to very high precision. The only solution is to treat their probability in a statistical sense, which we have already done in our treatment of wider binary systems \citep[][]{Kraus:2009uq}. We found that the probability of a chance alignment between two young association members is negligible for separations $<$10\arcsec, so we will proceed under the assumption that any pair of young stars constitutes a genuine binary system.

In Table 5, we list the mass ratios and component masses that we infer for the candidate companions brighter than $K=14$ and closer than 3\arcsec\, from their primary star, which we henceforth consider to be bound binary companions. For targets that were observed at multiple epochs, we list the independent estimates from each epoch. These properties were derived using the methods we describe in Section 3.3.

\section{Characterizing Multiplicity at the Bottom of the IMF}

The frequency and properties of multiple star systems offer important constraints of star formation processes, and the extreme disparity between the binary populations of the VLM population and higher-mass stars could provide a powerful test of star formation models. However, most of the large multiplicity surveys in the VLM regime have been conducted for old systems in the field. Constraints for young binary systems, especially those in dynamically primordial populations like Taurus and Upper Sco, are only now beginning to match the field surveys.

The archetypal concept of VLM multiplicity was established by a trio of high-resolution imaging surveys for nearby field targets. \citet[][]{Bouy:2003qc}, \citet[][]{Burgasser:2003mw}, and \citet[][]{Close:2003ud} all found that low-mass binaries are less common ($f\sim$10--15\% for separations of $\rho$$\ga$2--4 AU), more compact ($\rho$$\la$10--20 AU), and more symmetric ($q\ga$0.7--0.8, or $M_{sec}$$\sim$$M_{prim}$) than the corresponding population of solar-type binaries studied by \citet[][]{Duquennoy:1991zh}. The scarcity and tightly-bound nature of low-mass binaries led to suggestions that this indicated past strong dynamical interactions, perhaps consistent with the embryo ejection hypothesis for brown dwarf formation \citep[][]{Reipurth:2001ec}. However, the field population only allows an incomplete and muddled view of its primordial properties. The field represents a composite of many different formation environments, but it is probably dominated by stars formed in dense clusters \citep[e.g.,][]{Lada:2003qo}, so it is difficult to disentangle any environmental effects, especially those tied to primordial stellar density. Field brown dwarfs are also subject to a mass-age degeneracy, which makes it difficult to infer mass-dependent trends, and the steep mass-luminosity relation makes it difficult to identify companions which are much less massive than their primary stars \citep[][]{Chabrier:2000sh}.

These complications can be avoided by studying multiplicity in nearby star-forming regions and young associations. These populations have homogeneous and better-constrained initial conditions, their known age allows for a (model-dependent) resolution of the mass-age degeneracy, and their youth corresponds to a very shallow mass-luminosity relation that improves sensitivity to low-mass companions. Furthermore, these results can be directly compared to simulated stellar populations (e.g., Bate 2012), as we discuss in more detail in Section 5.3. The only tradeoff is that these populations are more distant than nearby field stars, imposing a resolution penalty against the discovery of binaries with small separations. Preliminary surveys have indicated that the field paradigm, with infrequent and tightly-bound binaries, is broadly consistent with several different formation environments \citep[][]{Kraus:2005qf,Kraus:2006ul,Konopacky:2007pd,Ahmic:2007hn}. However, they also indicated a further dependence of separation and frequency on mass within the VLM and substellar regime, and these mass-dependent effects can only be explored with a large binary survey among targets with known ages. 

\subsection{Bayesian Inference and Binary Population Statistics}

Binary population statistics are traditionally presented in terms of histograms of binary frequency versus separation or mass ratio, where the data is presented only for a range where the survey is complete. The analytic form of the preferred model is then fit to these histograms in order to infer the population properties. This approach has the virtue of simplicity, but estimating the probability density function (PDF) for the model's scale parameters is often difficult, especially if there are covariances between parameters. This method also is manifestly inadequate for handling heterogeneous datasets. If different stars have different detection limits, such as from being observed with different methods or under different atmospheric conditions, then simple histograms can be constructed only by appealing to completeness corrections that are themselves poorly constrained.

A better solution for working with heterogeneous data is to adopt a Bayesian approach, where the scale parameters of the model are assigned a prior PDF and that PDF is modified by each observation. This method exploits Bayes' theorem:

\begin{equation}
 P(\theta|O) \propto P(O|\theta) P(\theta)
\end{equation}

\noindent where $\theta$ represents the ``model'' (a set of scale parameters describing the functional form), $O$ represents the observation, $P(\theta|O)$ is the posterior PDF for the model (as a function of its parameters) given the data, $P(O|\theta)$ is the probability of obtaining an observation as a function of the model parameters, and $P(\theta)$ is the prior PDF for the model (again, as a function of its parameters). In cases with multiple observations (such as a survey of many targets), the posterior function for one observation is then used as the prior function for the next observation.

\citet[][]{Allen:2007rv} (hereafter A07) developed the relevant techniques for applying Bayesian statistics to VLM multiplicity, and we have largely adopted his approach in this work and a parallel survey of solar-type stars in Taurus \citep[][]{Kraus:2011tg}. We specifically describe the binary population in terms of a binary fraction $F$, a power-law mass ratio distribution with exponent $\gamma$, and a log-normal separation distribution with mean $\log(s)$ and standard deviation $\sigma$$_{\log(s)}$. We have adopted the same Poisson likelihood function as A07, but we will use a moderately different set of prior distributions. As for A07, we use constant priors for $\gamma$ and $\log(s)$. However, A07 seem to state that scale invariance requires the optimal unbiased prior for $\sigma_{\log(s)}$ to be proportional to $1/\sigma_{\log(s)}$, whereas it actually should be proportional to $1/\sigma_s$, or constant in $\sigma_{\log(s)}$.

The difference for $F$ is more subtle. A07 used a prior proportional to $1/F$, which was suggested to be suitable for a Poissonian variable. As we discussed in Kraus et al. (2011), the Poisson case is sometimes appropriate if there are many high-order multiples, though our newest analysis actually uses the Jeffries prior that is proportional to $1/\sqrt{F}$ (Kraus et al., in prep). However, binary companions exclude other binary companions in similar orbits (spanning at least a decade of semi major axis), so they are are not genuinely Poissonian. As we show below, ultra cool binaries typically only have small projected separations, perhaps not spanning much more than a decade of semi major axis, so the exclusionary effect should be quite significant. We are attempting to use completely uninformed priors for this analysis, but a more sophisticated prior that declines more quickly than $1/\sqrt{F}$ should considered for future analyses.

Another significant difference between our analysis and that of A07 is that we directly model the projected separation distribution, whereas A07 used the semimajor axis as a model parameter and then extrapolated a projected separation distribution using an assumed eccentricity distribution and randomly-distributed inclinations and phase angles. The eccentricity distribution for binary systems is still largely unconstrained for the separations and mass ranges that we consider, but most likely eccentricity distributions yield a separation distribution that is directly proportional to the semimajor axis distribution. As such, our results can be directly related to theoretical semimajor axis distributions once the eccentricity distribution is predicted by theory or measured by future surveys. We have also omitted the volume-completeness correction used by A07 to compensate for the overluminosity of similar-brightness binaries. The discovery surveys for most of our sample members were spatially limited, not flux- or volume-limited, so binary systems were equally likely to be detected. The high-resolution imaging techniques used in past surveys were themselves flux-limited, but we chose our LGSAO sample in part to compensate for this limit, so it should not significantly influence our results to invoke detections and detection limits from those past surveys where needed.

Our specific implementation of Bayesian analysis follows that of A07, defining two-dimensional functions of projected separation $\log(s)$ and mass ratio $q$ that denote the number of observations sensitive to each set of $\log(s)$ and $q$ (the ``window function'') and the corresponding number of companions with that set of parameters. We iterated our calculation over all mass ratios from 0 to 1 in steps of 0.01 and over all values of $\log(s)$ between 0.5 and 3.6 dex in steps of 0.1 dex; this range of $\log(s)$ was chosen to encompass the full range of projected (physics) separations for which imaging observations are sensitive, from 3 AU (due to the minimum resolution limit of the most sensitive surveys; Kraus et al. 2005, 2006) to 4500 AU (the maximum projected separation for which we have eliminated background star interlopers for most targets; Kraus \& Hillenbrand 2007, 2009).

Our own sample comprises the vast majority of available measurements in the VLM regime ($M\la$0.15 $M_{\sun}$), but almost all of the measurements for higher-mass stars must be adopted from previous surveys. As we describe in Appendix B and list in Table 7, we have specifically adopted the detections and detection limits for all stars with $M\la$0.5 $M_{\sun}$ for previous surveys of Taurus-Auriga \citep[][]{Ghez:1993xh,Simon:1995yi,Kraus:2006ul,Konopacky:2007pd}, Upper Sco \citep[][]{Kohler:2000lo,Kraus:2005qf,Biller:2011mw}, and Cha-I \citep[][]{Ahmic:2007hn,Lafreniere:2008fq}. In each case, we converted the measured angular separations (in mas) and flux ratios (in $\Delta$$m$) for their detections and detection limits into physical quantities (separations in AU and mass ratios) using the same methods that we applied to our own sample. This ensures a uniform set of inputs for our analysis, whereas each survey's inferred system properties were derived using different combinations of association distances/ages and pre-main sequence stellar evolutionary models.

Past studies have indicated that multiple star formation might be a mass-dependent process, so we have divided our sample into several bins. The stellar/substellar boundary represents a natural breaking point since it corresponds roughly to the M/L boundary for field objects, allowing for natural comparison to field samples. As we have indicated in our studies of solar-type multiplicity and wide multiplicity, systems with primary masses of $\ga$0.5 $M_{\sun}$ tend to have fundamentally different binary parameters, featuring a log-flat separation distribution and many very wide systems, so we have adopted this limit as the maximum for consideration in our sample. The desire for similar number statistics per bin therefore dictated four mass bins of similar width: $<$0.07 $M_{\sun}$, 0.07--0.15 $M_{\sun}$, 0.15--0.30 $M_{\sun}$, and 0.30--0.50 $M_{\sun}$. A more rigorous treatment might incorporate a mass dependence directly into our fit parameters ($F$, $\gamma$, $\log(s)$, and $\sigma$$_{\log(s)}$), but in the absence of any theoretical guidance on the functional form of this mass dependence, we will defer such analysis to a future study.

Our Bayesian analysis yields a PDF for all possible ``models'' that is defined across four dimensions, so we can not present the full results in a two-dimensional medium. However, any two parameters for which the covariance is small can be presented separately without discarding information. This independence allows us to instead present the results as a series of lower-dimensional surfaces, where PDF is integrated across the uncorrelated parameters in order to flatten its dimensionality. As we describe in the next section, our results can be described with a manageable number of 2D or 1D surfaces.

Finally, it is impossible to define a true PDF with only null detections, so we can not use this analysis for the lowest-mass bin ($M<0.07$ $M_{\sun}$) since it includes no resolved binary systems. Our choice to use generic conjugate priors does not disallow arbitrarily extreme values, such as small mean separations or steep power laws. If we do not have enough constraints (i.e. detected binaries) to force the PDF to zero at all extrema of the scale parameters, then the integrated probability will diverge and render the PDF unnormalizeable. Since we can not estimate a well-defined probability for any particular set of parameters being ``correct'', we instead must settle for a weaker result: the probability that a given set of parameters would have yielded our null detection. This measurement is equally valid for ruling out parameter space, but does not carry any explicitly affirmative value; regions where the model is less improbable are not necessarily regions where the model is probable. The act of ``flattening'' the PDF to visualizable 1D or 2D figures is also not defined for this type of constraint, so in figures where the PDF for higher-mass bins is flattened, we will instead show a cross-section through the lowest-mass PDF where we adopt the field T dwarf parameters suggested by \citet[][]{Burgasser:2006yp}: $\log(s)$$\sim$0.6, $\sigma$$_{\log(s)}$$\sim$0.3, and $\gamma$$\sim$4.2.

\subsection{The Mass-Dependent Parameters of the Multiple Star Population} 

The four-dimensional posterior PDF for our Bayesian analysis can be flattened to present six two-dimensional probability surfaces and four one-dimensional probability curves, but to convey the useful conclusions, we only need surfaces for covarying parameters and curves for non-covarying parameters. For our results, we will present two probability surfaces ($F$ versus $\log(s)$ and $F$ versus $\sigma$$_{\log(s)}$) and one probability curve ($\gamma$).

In general, we found little covariance between $\gamma$ and any other parameter, which is largely a result of the shallow mass-luminosity relation for young stars and brown dwarfs; at a given separation, most observations are either unable to detect any companions or sensitive to companions with almost all mass ratios. There is significant covariance between the binary frequency $F$ and the two parameters in the separation distribution, $\log(s)$ and $\sigma$$_{\log(s)}$. This degeneracy results from the inner working angle for most of the input datasets ($\sim$5--10 AU) being of similar order as the mean separation, since our measurements are consistent with a range of binary frequencies as long as an appropriate fraction of the companions are ``hidden'' inside the detection limit with a smaller mean separation and correspondingly wider standard deviation. As we discuss further in Section 5.3, a comparison to similar constraints from simulated stellar populations (e.g., Bate 2012) also can provide a direct test of star formation models and indicate future directions for their improvement.
 
\subsubsection{Frequencies and Separation Distributions for Stellar Multiplicity}

In Figure 6, we show the PDFs for our three stellar-mass bins as projected onto the $F$-$\log(s)$ and $F$-$\sigma$$_{\log(s)}$ planes. In the two lower-mass bins (0.30--0.15 $M_{\sun}$ and 0.15--0.07 $M_{\sun}$), there is a significant degeneracy between the overall binary frequency and the mean separation, where a smaller mean separation is paired with a higher frequency. This degeneracy is unavoidable when fitting a normal distribution whose mean is near or outside the fitting region; the first derivative (i.e. the slope) of the distribution across the fitting region yields the standard deviation with little ambiguity, but distinguishing between the total amplitude of the curve and the distance (in standard deviations) to the mean requires measurements of both the number of measurements and the second derivative (i.e. the change in slope) across the fitting region. Measuring each successive derivative requires either more $S/N$ or a wider fit regime. The highest-mass bin also shows some degeneracy with separation, but not to the same extent since its mean separation is outside of the typical inner working angles for many of the input surveys.

The $F$-$\log(s)$ locus in the highest-mass bin (0.3--0.5 $M_{\sun}$) is clearly distinct from the loci of the two lower bins (0.30--0.15 $M_{\sun}$ and 0.15--0.07 $M_{\sun}$), as its 90\% confidence region does not overlap with the same regions for the other bins. This indicates that the mean separation and/or the binary frequency are significantly higher for 0.3--0.5 $M_{\sun}$ stars. The strong degeneracies seen for the lower-mass bins make it difficult to draw any strong conclusions, but it appears that the frequency and/or mean separation for binary systems declines from the 0.15--0.30 $M_{sun}$ regime to the 0.07--0.15 $M_{\sun}$ regime. As we discuss in Section 5.2.2, this decline seems to continue in the substellar bin, though our constraint on its relative magnitude in each parameter is even weaker since we only have a null detection for that mass bin. The $F$-$\sigma$$_{\log(s)}$ loci are not as easy to interpret; the middle locus is biased to a larger standard deviation, but this might be the result of having 4 probable binary systems with separations of $>$1000 AU; if some of these pairs of stars are actually chance alignments of unrelated association members, then removing them would reduce the standard deviation and the mean separation by a significant amount.

The frequency-separation degeneracy must be addressed before we can draw any stronger conclusions. The most direct solution would be to increase the number statistics in our existing program, yielding a better estimate of the high-order derivatives in the separation distribution. However, this endeavor would be very observationally expensive; simulations show that even doubling our sample would not decrease the length of the degenerate locus, only its width. A less direct solution would be to expand the range of separations over which the distribution is constrained, either by observing at higher resolution (sampling more of the core separation distribution) or by searching for spectroscopic binaries (constraining the other wing of the separation distribution). We suggest that an RV survey would be significantly cheaper since it can exploit the multiplexing of wide-field multi-object spectrographs, plus the separation-frequency degeneracy that results from an RV survey's outer working angle should be perpendicular to the degeneracy from imaging surveys' inner working angle. Such surveys are currently being pursued for nearby young populations like the sigma Ori cluster \citep[][]{Maxted:2008gl} and the ONC \citep[][]{Tobin:2009fo}, and their results could be modeled with similar Bayesian techniques in order to produce constraints analogous to those shown in Figure 6. However, such a modeling effort is beyond the scope of the current work.

Finally, although Bayesian inference allows us to estimate the most general limits on the binary population, the degeneracies in those limits make it difficult to straightforwardly grasp the differences in our subsamples. We address this by forward-modeling from our four-dimensional PDF back into the range of separations and mass ratios where our observations could detection companions around most of our targets. The net result of this extrapolation is to implement a minor correction for incompleteness, but rather than adopting one assumed form for the underlying distributions, we implicitly integrated the correction over all possible distributions, weighted by the probability for each distribution. To this end, we have integrated over the entire four-dimensional PDF of each mass subsample to extrapolate the binary frequency at separations of 8--5000 AU and spanning all mass ratios of $0<q<1$. We find that in order of declining mass, our three subsamples (0.5--0.3 $M_{\sun}$, 0.30--0.15 $M_{\sun}$, and 0.15--0.07 $M_{\sun}$) have binary frequencies of $50^{+10}_{-9}\%$, $29^{+7}_{-6}\%$, and $21^{+7}_{-6}\%$ in this range of parameter space.

 \begin{figure*}
 \epsscale{0.75}
 \plotone{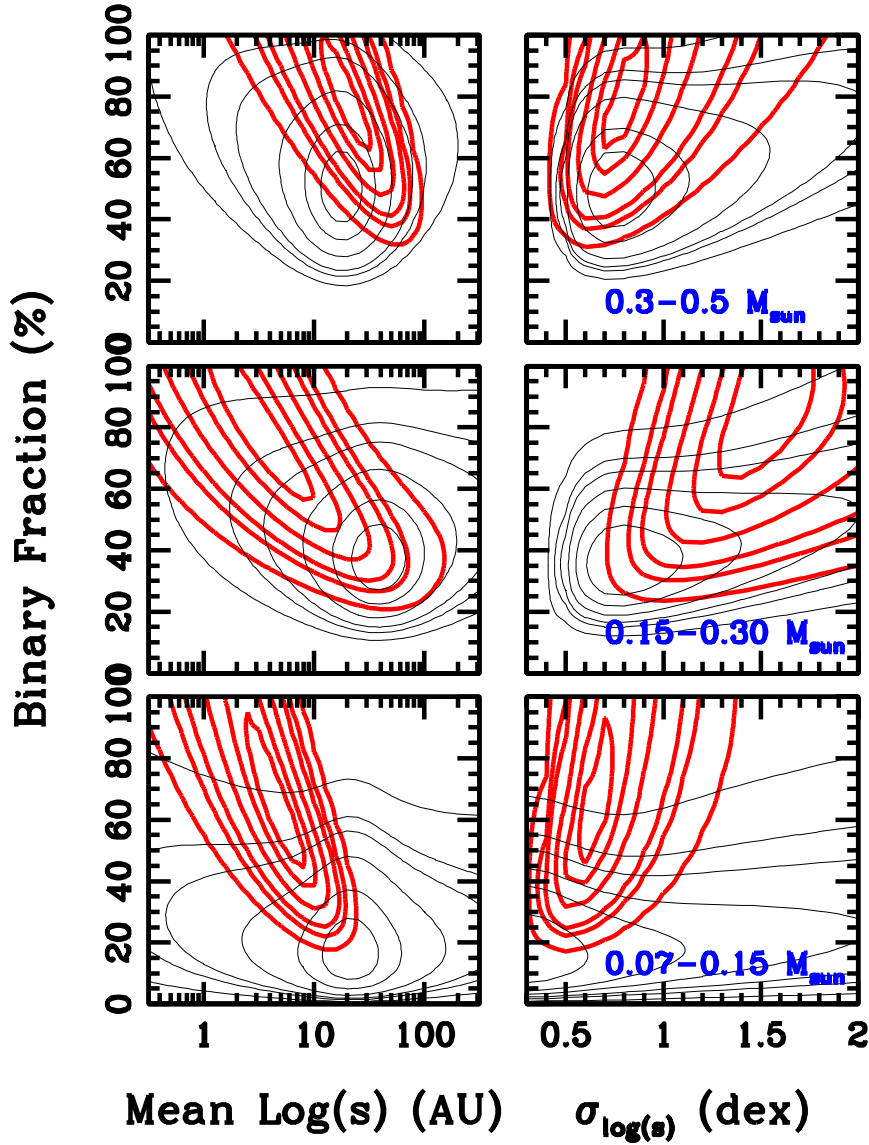}
 \caption{Posterior probability density functions for three mass ranges of stellar binaries. In each row, we plot the probability surface as projected onto the $F$-$\log(s)$ and $F$-$\sigma$$_{\log(s)}$ plane, showing contours that enclose total probability densities of 25\%, 50\%, 75\%, 90\%, 95\%, and 99\%. Red thick lines denote contours for our sample of observed binaries, while thin black lines denote contours for the synthetic binary population generated by Bate (2012) in a radiative SPH simulation (Section 5.3).}
 \end{figure*}

\subsubsection{Limits on Substellar Multiplicity}

As discussed above, we can not directly constrain the parameters of the substellar binary population because we did not discover any such binaries in our sample. However, we can estimate the probability of a null detection as a function of the four parameters in our model, ruling out a large portion of parameter space. In Figure 7, we show our null detection probability surfaces for the substellar mass bin in the $F$-$\log(s)$ and $F$-$\sigma$$_{\log(s)}$ planes. We can't integrate over the unplotted dimensions of our PDF since the integral diverges, so we instead show cross-sections for the most likely values as inferred by \citet[][]{Burgasser:2006yp}: $\log(s)$$\sim$0.6, $\sigma$$_{\log(s)}$$\sim$0.3 dex, and $\gamma$$\sim$4.2. We chose these parameters because the T dwarf sample studied by Burgasser et al. more closely matches our mass range than the full sample of MLT dwarfs studied by A07.

We find that for the given values of $\sigma$$_{\log(s)}$ and $\gamma$, we can not rule out any mean separations $\la$1 AU at $>$50\% confidence. However, we can rule out combinations of increasing mean separation and decreasing binary frequency; if the mean separation is 2 AU, then the binary frequency is $<$11\% at 50\% confidence and $<$38\% at 90\% confidence. If the mean separation is 4 AU, which is the maximum value consistent with the results of Burgasser et al., then the corresponding frequency limits are $<$4\% and $<$11\%, respectively. Conversely, if the total binary frequency is $\sim$20\%, then the 50\% and 90\% confidence limits on the mean separation are 1.6 AU and 2.8 AU. The corresponding probability surface for $\log(s)$$\sim$0.6 and $\gamma$$\sim$4.2 is strongly concentrated at low frequencies since this mean separation is very close to the inner working angle of our LGSAO survey, and therefore at least half of all companions should have been detectable. 

In summary, all of these limits for the substellar regime are extremely discrepant with respect to the confidence intervals for the higher-mass subsamples, which indicates that the the mass-dependent tightening of binary systems continues into the substellar regime. There are no well-defined and observationally-supported models for how low-mass binary systems form, so it is difficult to infer the underlying justification for the continued decline of system separations and/or frequencies into the substellar regime. However, the trend for declining separations and frequencies in the field appears to be established at very early ages and must result directly from the formation process.

 \begin{figure*}
 \epsscale{1.0}
 \plotone{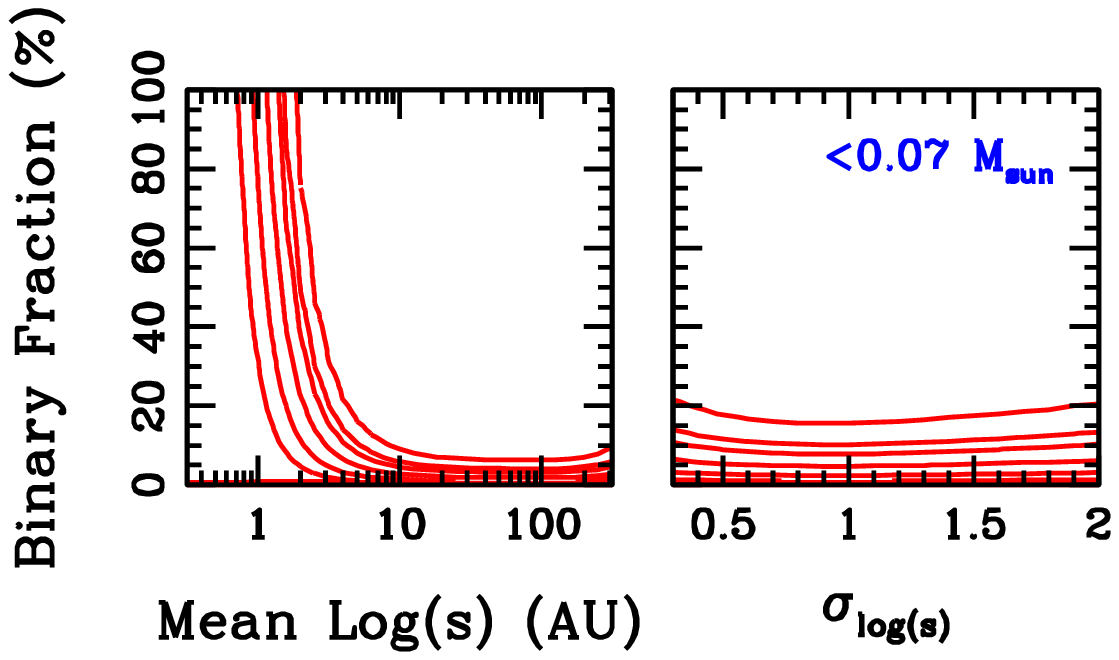}
 \caption{As in Figure 6, but for our substellar subsample. The posterior PDF can not be normalized, so we can not plot contours of enclosed total probability density or integrate across the unplotted parameters. We instead plot confidence contours on the probability surface for a null detection, and show cross-sections through the four-dimensional PDF at the most likely values inferred by the T dwarf multiplicity study of \citet[][]{Burgasser:2006yp}: $\log(s)$$\sim$0.6, $\sigma$$_{\log(s)}$$\sim$0.3 dex, and $\gamma$$\sim$4.2. We chose these parameters because the T dwarf sample studied by Burgasser et al. more closely matches our mass range than the full sample of MLT dwarfs studied by A07.}
 \end{figure*}

\subsubsection{Mass Ratio Distributions}

In Figure 8, we show the PDFs for our three stellar-mass bins as projected onto the $\gamma$ axis; we do not show any results for the substellar-mass bin because our null detection does not yield a useful constraint on its mass ratio distribution. Unlike for Figures 6 and 7, we decided to project the PDF onto an axis instead of a plane in order to display our constraints on the mass ratio distribution. There is no significant covariance between our constraints on $\gamma$ and those for other parameters, so this choice simplifies our presentation.

The 0.3--0.5 $M_{\sun}$ subsample has a best-fit slope of $\gamma$$=+0.18$$^{+0.33}_{-0.30}$, a value which is consistent with the linearly-flat mass ratio distribution found for higher-mass stars in young associations \citep[][]{Kraus:2008zr,Kraus:2011tg}. By contrast, the 0.15--0.30 $M_{\sun}$ subsample has a steeper slope of $\gamma$$=+1.02$$^{+0.59}_{-0.52}$, a value intermediate between the flat slope of higher-mass stars and the typically very steep power laws ($\gamma$$\sim$2--4) seen for late-M stars and L/T brown dwarfs in the field. 

Finally, the 0.07--0.15 $M_{\sun}$ subsample has a best-fit slope that is similar to the 0.15--0.30 $M_{\sun}$  bin, albeit with a very wide confidence interval, yielding $\gamma$$=0.96$$^{+0.70}_{-0.59}$. This result at first appears to contradict the overall trend for a steeping mass ratio distribution with declining mass that is seen in the field and would seem to lead from the 0.15--0.30 $M_{\sun}$ subsample. However, closer inspection of the sample suggests a possible solution. Of the 7 binary systems in the 0.07--0.15 $M_{\sun}$ bin with projected separations of $\la$20 AU, all have mass ratios of $\ga$0.5. By contrast, of the 5 binary systems with projected separations of $\ga$25 AU (GG Tau B, CFHT-Tau-17, and CFHT-Tau-18 in Taurus, Hn 13 in Cha-I, and RX J1558.1-2405b in Upper Sco),all but Hn 13 have a mass ratio of $\la$0.5. The corresponding limits on the mass ratio power-law exponent are $\gamma = 4.0^{+1.9}_{-1.6}$ for the closer subset and $\gamma = -0.3^{+0.7}_{-0.5}$ for the wider subset.

This trend is particularly intriguing because the five wider systems seem to approach or exceed the mass-maximum separation limit observed for field systems by \citet[][]{Burgasser:2003mw}, who observed that for VLM binary systems in the field, there is a mass-dependent upper envelope to binary system separations, $a_{max}\sim$1400$M_{tot}^2$. The five systems have typical total masses of $\sim$0.15 $M_{\sun}$, corresponding to maximum ``allowed'' separations of $\sim$30 AU. As such, they appear to unusual in both their separation and their mass ratio.

It is tempting to consider whether the markedly different mass function for wider VLM binary systems is a result of a different formation history. For example, wider binary systems most likely form earlier in the collapse of the progenitor molecular core. At these earlier stages, there is still more material left in the circumstellar envelope that might preferentially accrete onto the more massive binary component, driving the mass ratio further from unity. By contrast, close binary systems most likely form in the final stage of collapse, after much of the circumstellar envelope has been accreted into the central mass and little would remain for preferential accretion. Thus, if fragmentation tends to yield similar-mass components, then the epoch of fragmentation would dictate how far the mass ratio could evolve from unity. Since lower-mass binary systems also tend to have smaller separations, this would naturally lead to the trend for lower-mass binaries to have mass ratio distributions which are increasingly peaked at unity.

This model does not explain why these systems do not have analogs in the field, so we must appeal to a separate trend to justify their absence. Most of our sample is drawn from environments that are much less dense than typical star clusters; indeed, all of our targets are in loose associations that are unbound and should disperse within the next 10--50 Myr. By contrast, most stars form in denser clusters \citep[][]{Lada:2003qo} that are much more dynamically active and will ionize loosely-bound binary systems. For example, the separation distribution for solar-mass binary systems is truncated at separations of $\sim$300 AU in young clusters like the ONC \citep[][]{Kohler:2006wy} and at $\sim$100 AU in older clusters like Praesepe \citep[][]{Patience:2002qh}. VLM binary systems in Praesepe with equivalent binding energy would have separations a factor of $\sim$3 lower ($\sim$30 AU). Therefore, these systems might have counterparts in denser clusters, but those counterparts could be disrupted into their component singleton stars before reaching the field.

Finally, we must consider a more prosaic explanation as well. All of our targets are located at distances of $\sim$140 pc, so we can not resolve binary systems closer than $\sim$5--10 AU. It is possible that our ``unusually wide'' binary systems are actually hierarchical multiples, where one component of the wide pair appears fainter (and thus less massive) because it is actually a close double comprised of two stars that each contain approximately half the mass of the primary. This would yield a total mass ratio close to unity in the wide pair, plus the higher total mass would allow for a correspondingly wider separation without violating the $a_{max}$-$M_{tot}$ relation. Large field surveys are starting to uncover a significant number of the very rare systems that appear at first to violate this relation 
\citep[][]{Caballero:2007oe,Artigau:2007uk,Radigan:2009aj,Dhital:2010be}], but followup high-resolution imaging has shown that at least some of them are hierarchical triples or even quadruples \citep[][]{Law:2010vn}.

 \begin{figure}
 \plotone{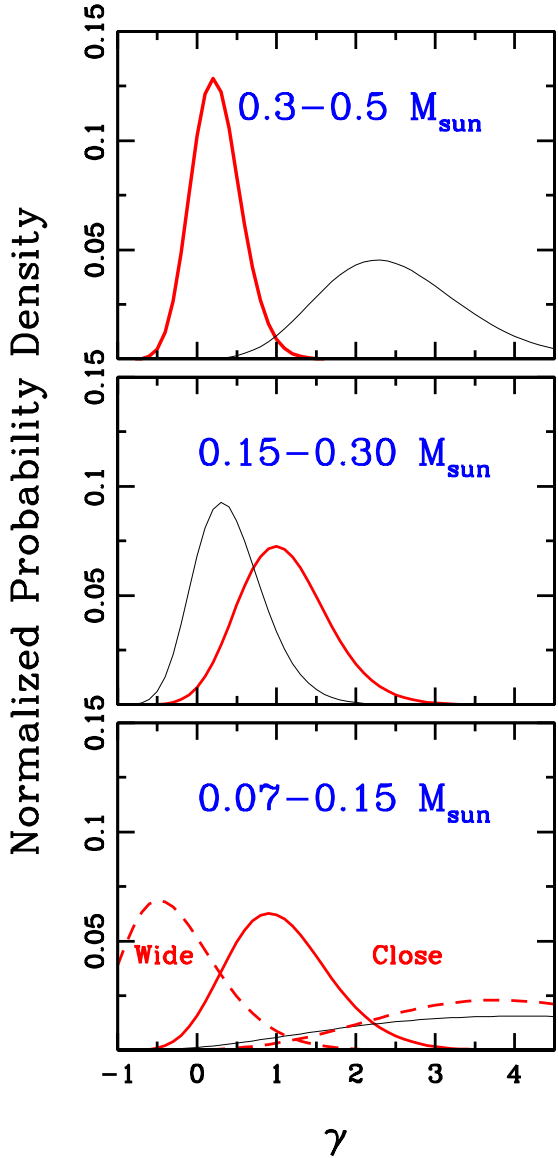}
 \caption{Posterior probability density functions for three mass ranges of stellar binaries in our sample. In each row, we plot the probability curve as projected onto the $\gamma$ axis, denoting our confidence interval on the power-law exponent for the mass ratio distribution. As we discuss in the text, we also show separate fits in the 0.07--0.15 $M_{\sun}$ subsample for systems with separations of $>$25 AU (5 systems) and $<$25 AU (7 systems); close binary systems have a mass ratio distribution that is strongly peaked at unity, while four of the five wider binary systems have mass ratios of $<$0.5. As for Figure 6, our observational results are shown with thick red lines, while the synthetic population of Bate (2012) is shown with thin black lines (Section 5.3). The two separation regimes in the 0.07--0.15 $M_{\sun}$ subsample are denoted with red dashed lines.}
 \end{figure}

\subsection{Comparison to Star Formation Models}

Smoothed particle hydrodynamic simulations are now capable of producing synthetic populations of young stars with statistically robust sizes (e.g., Bate 2009; Offner et al. 2010; Bate 2012), including samples of synthetic binaries that are equivalent in size to our observed samples of real systems. These synthetic binary populations provide a natural comparison sample for the results of our survey. The only synthetic population large enough to investigate mass-dependent trends was recently reported by Bate (2012), who simulated a cluster of $183$ stars and brown dwarfs, including $42$ multiple systems. Their simulation only ran for $2 \times 10^5$ yr (due to computational expense), and had to end before the cluster dispersed all of its gas or became unbound, so the synthetic population likely was still evolving at the conclusion. However, it still provides an illustrative comparison sample. To produce a direct comparison, we have subjected their binary sample to the same Bayesian formalism as our observed sample, with the caveats that we set no detection limits (since all binaries are known) and we treated the semimajor axis as equivalent to the projected separation (since they are statistically equal to within nearly unity; Dupuy et al. 2011). 

In Figures 6 and 8, we plot the confidence intervals corresponding to the posterior PDF as marginalized to the ($F$,$\mu_{\log(\rho)}$) and ($F$,$\sigma_{\log(\rho)}$) planes and the $\gamma$ line. These comparisons show that simulations reproduce a binary population that is broadly similar to that seen in observations; binaries have approximately the correct frequency, semimajor axes, and companion masses. This consistency is especially intriguing because the simulated environment is much denser than the regions we have observed, and hence dynamical interactions shape the synthetic population to a much greater degree. A more detailed comparison shows that not all features are consistent, though. 

The simulations reproduce the most distinctive feature of real binary populations: a binary frequency which declines with decreasing primary mass. However, the separation distribution does not show the corresponding trend toward smaller semimajor axes, with a mean separation of $\sim$20--40 AU across the entire mass range. Furthermore, the standard deviation of the separation distribution only declines in the least massive bin, where it denotes a paucity of binaries at both $>$100 AU (in agreement with observations) and $<$5 AU (in sharp contrast with observations). The mass ratio distribution also does not match the expected trend for lower-mass systems to have mass ratios near unity. The lowest-mass bin shows a very strong bias toward such systems, but the highest-mass bin also shows a similar bias. 

These discrepancies might be a result of not following the binary formation process to its conclusion. Accretion of additional circumstellar material should further modify the masses of the binary companions, and their semimajor axes also could be modified if the specific angular momentum of the accreted material is much higher or lower than for the binary components. Furthermore, dynamical interactions should disrupt or harden binary systems, and both effects would preferentially bias lower-mass binaries to have smaller semimajor axes than higher-mass binaries. The use of sink particles (with radius $r = 0.5$ AU) also could affect processes occurring on AU scales, inhibiting the simulated production of tight binary systems. 

If longer simulations can not improve the level of agreement, then it might suggest that other changes are required. The simulations now incorporate radiative feedback (Bate 2009; Bate 2012), and other simulations have suggested that magnetic fields are not likely to be significant on small scales (Price \& Bate 2009). One possible avenue is to simulate different initial conditions. The simulations of Bate (2012) currently begin with an isothemal sphere of uniform density, which does not match the configurations seen for pre-stellar environments like the Pipe Nebula (e.g., Lada et al. 2009). Another possible variable to change is the turbulent power spectrum, which might affect the properties of binary systems (e.g., Goodwin et al. 2004).

\section{Conclusions}

We have presented the results of a large-scale survey of multiplicity at the bottom of the IMF in several nearby young associations. We have confirmed the overall trend observed in the field for lower-mass binary systems to be less frequent and more compact, including a null detection for any substellar binary systems with separations wider than $\sim$5--10 AU. In the stellar-mass regime, we confirm that the binary frequency and binary separations decline between masses of 0.5 $M_{\sun}$ and 0.08 $M_{\sun}$, though a degeneracy between the binary frequency and the mean binary separation make it difficult to distinguish the degree of the decline in each parameter. We also confirm that the mass ratio distribution becomes progressively more concentrated at $q\sim$1 for declining masses. However, we also note that a small number of systems appear to have unusually wide separations and low mass ratios for their system mass; this could indicate a secondary channel for low-mass binary formation, though unresolved high-order multiplicity could explain the unusual nature of some systems. Finally, we compare our results to synthetic binary populations generated by SPH simulations, finding that while models now reproduce the mass-dependent frequency of multiple systems, differences remain in the mass-dependent separation and mass ratio distributions.

\section{Appendix A: A Model for Star Counts in the K Band}

The use of star count models was pioneered by \citet[][]{Bahcall:1980lr} in order to study the structure of the Milky Way. Their procedure invoked a simple two-component model of the galaxy (composed of a disk and a spheroid) to characterize the density of stars as a function of position in the galaxy. The integrated luminosity function along any sightline through this distribution would then yield the number of stars as a function of magnitude for that location on the celestial sphere. The model has since been updated to include two disk components, the thin and thick disks \citep[e.g.,][]{Gilmore:1983fj}, as well as separate components for the bulge and halo \citep[e.g.,][]{Jackson:2002kx}.

Bahcall \& Soneira originally used observational star counts in order to determine the scale heights and scale radii for each component of the galaxy. However, this process can also be inverted; given an adopted luminosity function and a set of scale heights and scale radii, it is possible to predict the number of stars per magnitude for any arbitrary position on the sky. We have developed an updated version of the Bahcall \& Soneira models in order to predict faint source counts in our K band observations, characterizing the rate of background star contamination.

We adopted our K-band luminosity functions from several sources in the literature. We directly invoked the well-known K-band luminosity function for field giants as described by \citet[][]{Mamon:1982vn}. The luminosity function for field dwarfs has only been measured in other filters, so we invoked the V-band luminosity function for A--K dwarfs from \citet[][]{Reid:2002sn}, the J-band luminosity function for M0--M6 dwarfs from \citet[][]{Reid:2003fr}, and the J-band luminosity function for M7--L8 dwarfs from \citet[][]{Cruz:2007ij}. In each case, we used the magnitude-SpT relations of \citet[][]{Kraus:2007mz} and the color-SpT relations of \citet[][]{Bessell:1988ri} to calculate the corresponding K-band luminosity function.

The scale parameters for Milky Way structural distributions, and even the functional forms themselves, have been updated numerous times since Bahcall \& Soneira derived their original estimates. We have chosen to characterize the two disk components using exponential scale heights and scale radii and the halo using a power-law scale exponent and an oblate axis ratio. We did not fit the bulge because its triaxial distribution is still somewhat uncertain and because all of our targets are $>$20$^o$ from the Galactic Center. Thus, the resulting functional form is:

\[
\rho(R,Z)=\rho(R_{\sun},0)\times\lgroup
\exp\left(\frac{R_{\sun}-R}{L_{thin}}-\frac{Z}{H_{thin}}\right)+
\]
\[
f_{thick}\exp\left(\frac{R_{\sun}-R}{L_{thick}}-\frac{Z}{H_{thick}}\right)+
\]
\[
f_{halo}\left(\frac{R_{\sun}}{\sqrt{R^2+(Z/q_{halo})^2}}\right)^{n_{halo}}\rgroup
\]

\noindent where $R$ and $Z$ are cylindrical Galactocentric coordinates, $R_{\sun}$ is the solar Galactocentric radius, $f_x$ denotes the normalized density of each component in the solar neighborhood (relative to the thin disk), $L_x$ denotes a scale radius, $H_x$ denotes a scale height, $q_{halo}$ is the oblate axis ratio for the halo, $n_{halo}$ is the power law exponent for the halo, and $\rho(R_{\sun},0)$ is the present-day mass function in the solar neighborhood.

The parameters for the disks and halo were estimated most recently by \citet[][]{Juric:2008qe}, using positions and photometric distances for stars from the Sloan Digital Sky Survey to directly fit the three-dimensional distributions of stars. Based on the distribution of disk M dwarfs, they found that the two disk components have scale heights of $H_{thin}=300$ pc and $H_{thick}=900$ pc and corresponding scale radii of $L_{thin}=2600$ pc and $L_{thick}=3600$ pc; the normalized density of thick disk stars in the solar neighborhood is $f_{thick}=0.12$. Based on the distribution of main-sequence turnoff stars, they found that the halo has an axis ratio of $q_{halo}=0.64$, a radial power-law exponent of $n_{halo}=-2.8$, and a normalized local density of $f_{halo}=5\times10^{-3}$.

However, we found from comparisons to 2MASS that the parameters of \citet[][]{Juric:2008qe} yielded a radial gradient in thin disk density that was too steep, overestimating the density of thin disk stars toward the Galactic center and underestimating the density toward the Galactic anticenter. Based on observations at very high galactic latitudes, we found that their parameters also overestimated the number of thick disk stars. Older studies \citep[e.g.,][]{Chen:2001qy,Siegel:2002zr} have found that a larger thin disk scale radius (similar to the thick disk, $\sim$3600 pc) and a lower fraction of thick disk stars in the solar neighborhood ($\sim$0.06) produce acceptable fits for other datasets. These values also fit our data, so we have adopted them instead.

Finally, we accounted for dust obscuration by assuming that the dust density approximately traces the thin disk; this result is roughly consistent with observations of nearby edge-on disk galaxies \citep[e.g.,][]{Bianchi:2007fk}. The total integrated extinction along a sightline (in magnitudes) is then proportional to the total integrated dust density. We normalized the extinction by assuming that dust causes one magnitude of V band extinction per kiloparsec in the Solar neighborhood, (0.11 magnitude of K band extinction, based on the reddening law of \citet[][]{Schlegel:1998yj}). All of our science targets are at intermediate galactic latitudes (15$^o$$<|b|<$30$^o$), so the total effect is $\la$0.5 mag in all cases.

We did not include the effect of residual molecular cloud material around our science targets because extinction measurements from the literature only include foreground obscuration, not background obscuration. The IRAS extinction maps of \citet[][]{Schlegel:1998yj} would provide rough estimates, but the obscuring material is usually patchy on scales smaller than the IRAS resolution, so any correction would be very uncertain. We prefer to overestimate source counts for all targets (a conservative error) rather than to risk underestimating source counts for some. This problem does not affect most Upper Sco members because its natal gas and dust has already dispersed.
  
In Figure 9, we plot the predicted K band source counts as a function of magnitude for four sightlines that correspond to nearby stellar populations. We find that the integrated density of all stars brighter than $K=20$ varies quite significantly, from 48 arcmin$^{-2}$ on the eastern edge of Upper Sco to 1.2 arcmin$^{-2}$ in the middle of Coma Berenices. We also show the observed 2MASS source counts for a 1$^o$ field surrounding each sight line; in all cases, our predictions agree with 2MASS predictions down to its 10$\sigma$ detection limit ($K=14.3$). Finally, we also show the K band galaxy source counts as determined from numerous extragalactic surveys \citep[][]{Cimatti:2002kx}. Galaxies only contribute significantly in Upper Sco at $K\ga$21, but they are a significant source of background contamination for lower-density fields (e.g. $K\ga$17 in Taurus).

 \begin{figure*}
 \epsscale{1.0}
 \plotone{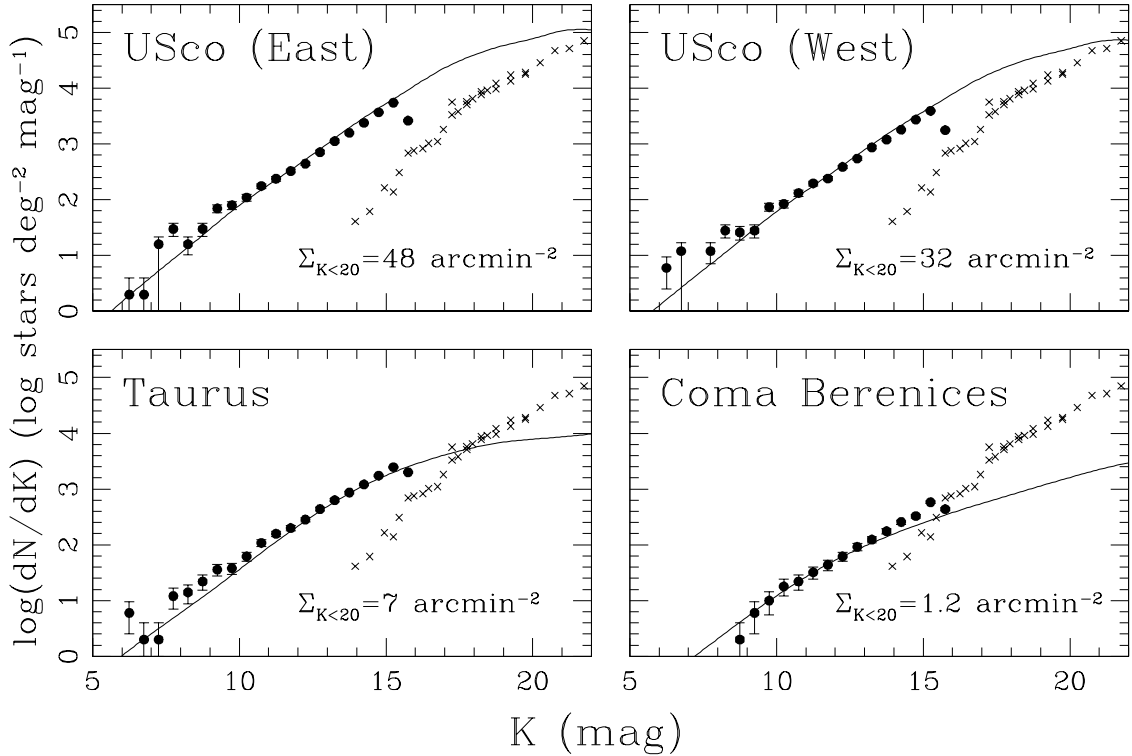}
 \caption{K-band source counts for four sightlines corresponding to nearby clusters or associations: eastern Upper Sco (16:00:00, -22:00:00), western Upper Sco (16:20:00, -22:00:00), Taurus (4:30:00, +25:00:00), and Coma Berenices (12:30:00, +26:00:00). We include Coma Berenices specifically because it is located at the galactic pole, sampling a very different sightline through the Milky Way. The solid line shows the predicted source counts from our model, filled circles show the empirical source counts for that sightline from 2MASS, and crosses show galaxy source counts as summarized by Cimatti et al. (2003). Our model shows excellent agreement with 2MASS; the empirical source counts diverge at faint magnitudes for Coma Ber because background galaxies dominate over Milky Way stars at $K\ga$15.}
 \end{figure*}

\section{Appendix B: Results from Previous Multiplicity Surveys}

The past two decades have seen numerous multiplicity surveys of young low-mass stars and brown dwarfs; most featured sample sizes of $N\sim$10-50, but in aggregate, they span several hundred targets. As we show in Section 5, a compilation of all such surveys can offer far better limits than any one survey on its own. In this appendix, we assemble and parse a large set of previous multiplicity surveys in Taurus, Upper Sco, and Cha-I in order to include their results in our Bayesian analysis.

Our composite sample draws from all of the surveys that reported their null detections and their detection limits as a function of separation. In Cha-I, we include the two surveys by \citet[][]{Ahmic:2007hn} and \citet[][]{Lafreniere:2008fq}. In Taurus, we include the seven surveys by \citet[][]{Ghez:1993xh} \citet[][]{Simon:1995yi}, \citet[][]{Sartoretti:1998bd}, \citet[][]{Kraus:2006ul}, \citet[][]{Konopacky:2007pd}, and \citet[][]{Duchene:2007uq}. In Upper Sco, we draw on the two surveys by \citet[][]{Kohler:2000lo} and \citet[][]{Kraus:2005qf}. Finally, for all regions, we draw on our past results for wide binary systems \citep[][]{Kraus:2007ve,Kraus:2009uq}.

Several other large surveys have been conducted, but we can not use their results in our Bayesian analysis because they either do not report their detection limits or do not list their null detections. We also omitted the results of our aperture masking surveys for both Taurus and Upper Sco \citep[][]{Kraus:2008zr,Kraus:2011tg} because they did not achieve significant completeness for any targets with $M_{prim} < 0.5$ $M_{\sun}$. However, in cases where those surveys did identify a new binary companion, we amended the assumed ``primary'' mass used in our analysis to reflect the contribution of that companion.

The construction of our sample is also complicated by the need to realistically consider hierarchical multiple systems. There are several such systems with primary masses of $M_{prim}$$<$0.5 $M_{\sun}$, and all of the binary pairs also have the potential to host additional components. Our solution includes several steps. First, we adopt the detection limits for a given primary star out to 3 times the projected separation $\rho$ to its binary companion; we chose this limit because the minimum stable ratio of semimajor axes for a hierarchical triple is $\sim$2--3 \citep[][]{Szebehely:1977mz}, and on a statistical basis, the projected separations for a sample of binary systems are similar to their semimajor axes \citep[$\rho \sim 1.26 a$;][]{Fischer:1992if}. Second, in considering limits for additional companions at $>$3$\rho$, we sum the masses of the inner binary pair and treat it as a single, more massive primary. Finally, we only consider detection limits for binary secondaries out to projected separations of $1/3$ the projected separation to the system primary, and only if the binary secondary is separated from its primary by $>$6\arcsec\, \citep[$\sim$1000 AU, the typical scale of a protostellar disk+envelope system;][]{Enoch:2009mp} since otherwise the existence of the primary star might have influenced the binary secondary's subsequent collapse and potential for fragmentation.

In Table 6, we list all of the known binary systems that we include in our Bayesian analysis, complementing the systems that we list in Tables 3 and 5. In Table 7, we list the corresponding mass ratio detection limits for all targets from our survey and from the literature. Each survey reported its detection limits at different points in the separation-mass ratio curve, so we have linearly interpolated between those surveys' listed values in order to produce a uniform grid of limits.

\acknowledgements

We thank Russel White for many insightful discussions of multiple star formation that have helped to shape our conclusions, as well as the anonymous referee for providing a detailed and very helpful critique of this work. We also thank the Keck LGSAO team for their efforts in developing and supporting a valuable addition to the observatory. ALK was supported by a NASA Origins grant to LAH, by a SIM Science Study, and by NASA through Hubble Fellowship grant 51257.01 awarded by STScI, which is operated by AURA, Inc. for NASA, under contract NAS 5-26555. This work makes use of data products from 2MASS, which is a joint project of the University of Massachusetts and the IPAC/Caltech, funded by NASA and the NSF.

The observations presented herein were obtained at the W.M. Keck Observatory, which is operated as a scientific partnership between Caltech, the University of California, and NASA. The observatory was made possible by the generous financial support of the W.M. Keck Foundation.The authors also wish to recognize and acknowledge the very significant cultural role and reverence that the summit of Mauna Kea has always had within the indigenous Hawaiian community. We are most fortunate to have the opportunity to conduct observations from this mountain.


\bibliographystyle{apj.bst}
\bibliography{krausbib}

\begin{thebibliography}{82}
\expandafter\ifx\csname natexlab\endcsname\relax\def\natexlab#1{#1}\fi

\bibitem[{{Ahmic} {et~al.}(2007){Ahmic}, {Jayawardhana}, {Brandeker}, {Scholz},
  {van Kerkwijk}, {Delgado-Donate}, \& {Froebrich}}]{Ahmic:2007hn}
{Ahmic}, M., {Jayawardhana}, R., {Brandeker}, A., {Scholz}, A., {van Kerkwijk},
  M.~H., {Delgado-Donate}, E., \& {Froebrich}, D. 2007, \apj, 671, 2074

\bibitem[{{Allen}(2007)}]{Allen:2007rv}
{Allen}, P.~R. 2007, \apj, 668, 492

\bibitem[{{Artigau} {et~al.}(2007){Artigau}, {Lafreni{\`e}re}, {Doyon},
  {Albert}, {Nadeau}, \& {Robert}}]{Artigau:2007uk}
{Artigau}, {\'E}., {Lafreni{\`e}re}, D., {Doyon}, R., {Albert}, L., {Nadeau},
  D., \& {Robert}, J. 2007, \apjl, 659, L49

\bibitem[{{Bahcall} \& {Soneira}(1980)}]{Bahcall:1980lr}
{Bahcall}, J.~N., \& {Soneira}, R.~M. 1980, \apjs, 44, 73

\bibitem[{{Baraffe} {et~al.}(1998){Baraffe}, {Chabrier}, {Allard}, \&
  {Hauschildt}}]{Baraffe:1998yo}
{Baraffe}, I., {Chabrier}, G., {Allard}, F., \& {Hauschildt}, P.~H. 1998, \aap,
  337, 403

\bibitem[{{Bate} \& {Bonnell}(1997)}]{Bate:1997kx}
{Bate}, M.~R., \& {Bonnell}, I.~A. 1997, \mnras, 285, 33

\bibitem[{{Bessell} \& {Brett}(1988)}]{Bessell:1988ri}
{Bessell}, M.~S., \& {Brett}, J.~M. 1988, \pasp, 100, 1134

\bibitem[{{Bianchi}(2007)}]{Bianchi:2007fk}
{Bianchi}, S. 2007, \aap, 471, 765

\bibitem[{{Biller} {et~al.}(2011){Biller}, {Allers}, {Liu}, {Close}, \&
  {Dupuy}}]{Biller:2011mw}
{Biller}, B., {Allers}, K., {Liu}, M., {Close}, L.~M., \& {Dupuy}, T. 2011,
  \apj, 730, 39

\bibitem[{{Bouy} {et~al.}(2003){Bouy}, {Brandner}, {Mart{\'{\i}}n}, {Delfosse},
  {Allard}, \& {Basri}}]{Bouy:2003qc}
{Bouy}, H., {Brandner}, W., {Mart{\'{\i}}n}, E.~L., {Delfosse}, X., {Allard},
  F., \& {Basri}, G. 2003, \aj, 126, 1526

\bibitem[{{Brice{\~n}o} {et~al.}(2002){Brice{\~n}o}, {Luhman}, {Hartmann},
  {Stauffer}, \& {Kirkpatrick}}]{Briceno:2002jo}
{Brice{\~n}o}, C., {Luhman}, K.~L., {Hartmann}, L., {Stauffer}, J.~R., \&
  {Kirkpatrick}, J.~D. 2002, \apj, 580, 317

\bibitem[{{Burgasser} {et~al.}(2006){Burgasser}, {Kirkpatrick}, {Cruz}, {Reid},
  {Leggett}, {Liebert}, {Burrows}, \& {Brown}}]{Burgasser:2006yp}
{Burgasser}, A.~J., {Kirkpatrick}, J.~D., {Cruz}, K.~L., {Reid}, I.~N.,
  {Leggett}, S.~K., {Liebert}, J., {Burrows}, A., \& {Brown}, M.~E. 2006,
  \apjs, 166, 585

\bibitem[{{Burgasser} {et~al.}(2003){Burgasser}, {Kirkpatrick}, {Reid},
  {Brown}, {Miskey}, \& {Gizis}}]{Burgasser:2003mw}
{Burgasser}, A.~J., {Kirkpatrick}, J.~D., {Reid}, I.~N., {Brown}, M.~E.,
  {Miskey}, C.~L., \& {Gizis}, J.~E. 2003, \apj, 586, 512

\bibitem[{{Caballero}(2007)}]{Caballero:2007oe}
{Caballero}, J.~A. 2007, \aap, 462, L61

\bibitem[{{Chabrier} {et~al.}(2000){Chabrier}, {Baraffe}, {Allard}, \&
  {Hauschildt}}]{Chabrier:2000sh}
{Chabrier}, G., {Baraffe}, I., {Allard}, F., \& {Hauschildt}, P. 2000, \apj,
  542, 464

\bibitem[{{Chauvin} {et~al.}(2004){Chauvin}, {Lagrange}, {Dumas}, {Zuckerman},
  {Mouillet}, {Song}, {Beuzit}, \& {Lowrance}}]{Chauvin:2004bd}
{Chauvin}, G., {Lagrange}, A., {Dumas}, C., {Zuckerman}, B., {Mouillet}, D.,
  {Song}, I., {Beuzit}, J., \& {Lowrance}, P. 2004, \aap, 425, L29

\bibitem[{{Chen} {et~al.}(2001){Chen}, {Stoughton}, {Smith}, {Uomoto}, {Pier},
  {Yanny}, {Ivezi{\'c}}, {York}, {Anderson}, {Annis}, {Brinkmann}, {Csabai},
  {Fukugita}, {Hindsley}, {Lupton}, {Munn}, \& {the SDSS
  Collaboration}}]{Chen:2001qy}
{Chen}, B., {et~al.} 2001, \apj, 553, 184

\bibitem[{{Cimatti} {et~al.}(2002){Cimatti}, {Mignoli}, {Daddi}, {Pozzetti},
  {Fontana}, {Saracco}, {Poli}, {Renzini}, {Zamorani}, {Broadhurst},
  {Cristiani}, {D'Odorico}, {Giallongo}, {Gilmozzi}, \&
  {Menci}}]{Cimatti:2002kx}
{Cimatti}, A., {et~al.} 2002, \aap, 392, 395

\bibitem[{{Close} {et~al.}(2003){Close}, {Siegler}, {Freed}, \&
  {Biller}}]{Close:2003ud}
{Close}, L.~M., {Siegler}, N., {Freed}, M., \& {Biller}, B. 2003, \apj, 587,
  407

\bibitem[{{Correia} {et~al.}(2006){Correia}, {Zinnecker}, {Ratzka}, \&
  {Sterzik}}]{Correia:2006pi}
{Correia}, S., {Zinnecker}, H., {Ratzka}, T., \& {Sterzik}, M.~F. 2006, \aap,
  459, 909

\bibitem[{{Cruz} {et~al.}(2007){Cruz}, {Reid}, {Kirkpatrick}, {Burgasser},
  {Liebert}, {Solomon}, {Schmidt}, {Allen}, {Hawley}, \& {Covey}}]{Cruz:2007ij}
{Cruz}, K.~L., {et~al.} 2007, \aj, 133, 439

\bibitem[{{de Zeeuw} {et~al.}(1999){de Zeeuw}, {Hoogerwerf}, {de Bruijne},
  {Brown}, \& {Blaauw}}]{de-Zeeuw:1999xv}
{de Zeeuw}, P.~T., {Hoogerwerf}, R., {de Bruijne}, J.~H.~J., {Brown}, A.~G.~A.,
  \& {Blaauw}, A. 1999, \aj, 117, 354

\bibitem[{{Delgado-Donate} {et~al.}(2004){Delgado-Donate}, {Clarke}, {Bate}, \&
  {Hodgkin}}]{Delgado-Donate:2004id}
{Delgado-Donate}, E.~J., {Clarke}, C.~J., {Bate}, M.~R., \& {Hodgkin}, S.~T.
  2004, \mnras, 351, 617

\bibitem[{{Dhital} {et~al.}(2010){Dhital}, {West}, {Stassun}, \&
  {Bochanski}}]{Dhital:2010be}
{Dhital}, S., {West}, A.~A., {Stassun}, K.~G., \& {Bochanski}, J.~J. 2010, \aj,
  139, 2566

\bibitem[{{Duch{\^e}ne} {et~al.}(2007){Duch{\^e}ne}, {Bontemps}, {Bouvier},
  {Andr{\'e}}, {Djupvik}, \& {Ghez}}]{Duchene:2007uq}
{Duch{\^e}ne}, G., {Bontemps}, S., {Bouvier}, J., {Andr{\'e}}, P., {Djupvik},
  A.~A., \& {Ghez}, A.~M. 2007, \aap, 476, 229

\bibitem[{{Duquennoy} \& {Mayor}(1991)}]{Duquennoy:1991zh}
{Duquennoy}, A., \& {Mayor}, M. 1991, \aap, 248, 485

\bibitem[{{Enoch} {et~al.}(2009){Enoch}, {Evans}, {Sargent}, \&
  {Glenn}}]{Enoch:2009mp}
{Enoch}, M.~L., {Evans}, N.~J., {Sargent}, A.~I., \& {Glenn}, J. 2009, \apj,
  692, 973

\bibitem[{{Fischer} \& {Marcy}(1992)}]{Fischer:1992if}
{Fischer}, D.~A., \& {Marcy}, G.~W. 1992, \apj, 396, 178

\bibitem[{{Ghez} {et~al.}(1993){Ghez}, {Neugebauer}, \&
  {Matthews}}]{Ghez:1993xh}
{Ghez}, A.~M., {Neugebauer}, G., \& {Matthews}, K. 1993, \aj, 106, 2005

\bibitem[{{Ghez} {et~al.}(2008){Ghez}, {Salim}, {Weinberg}, {Lu}, {Do}, {Dunn},
  {Matthews}, {Morris}, {Yelda}, {Becklin}, {Kremenek}, {Milosavljevic}, \&
  {Naiman}}]{Ghez:2008my}
{Ghez}, A.~M., {et~al.} 2008, \apj, 689, 1044

\bibitem[{{Gilmore} \& {Reid}(1983)}]{Gilmore:1983fj}
{Gilmore}, G., \& {Reid}, N. 1983, \mnras, 202, 1025

\bibitem[{{Hillenbrand} \& {White}(2004)}]{Hillenbrand:2004bh}
{Hillenbrand}, L.~A., \& {White}, R.~J. 2004, \apj, 604, 741

\bibitem[{{Jackson} {et~al.}(2002){Jackson}, {Ivezi{\'c}}, \&
  {Knapp}}]{Jackson:2002kx}
{Jackson}, T., {Ivezi{\'c}}, {\v Z}., \& {Knapp}, G.~R. 2002, \mnras, 337, 749

\bibitem[{{Juri{\'c}} {et~al.}(2008){Juri{\'c}}, {Ivezi{\'c}}, {Brooks},
  {Lupton}, {Schlegel}, {Finkbeiner}, {Padmanabhan}, {Bond}, {Sesar},
  {Rockosi}, {Knapp}, {Gunn}, {Sumi}, {Schneider}, {Barentine}, {Brewington},
  {Brinkmann}, {Fukugita}, {Harvanek}, {Kleinman}, {Krzesinski}, {Long},
  {Neilsen}, {Nitta}, {Snedden}, \& {York}}]{Juric:2008qe}
{Juri{\'c}}, M., {et~al.} 2008, \apj, 673, 864

\bibitem[{{K{\"o}hler} {et~al.}(2000){K{\"o}hler}, {Kunkel}, {Leinert}, \&
  {Zinnecker}}]{Kohler:2000lo}
{K{\"o}hler}, R., {Kunkel}, M., {Leinert}, C., \& {Zinnecker}, H. 2000, \aap,
  356, 541

\bibitem[{{K{\"o}hler} {et~al.}(2006){K{\"o}hler}, {Petr-Gotzens},
  {McCaughrean}, {Bouvier}, {Duch{\^e}ne}, {Quirrenbach}, \&
  {Zinnecker}}]{Kohler:2006wy}
{K{\"o}hler}, R., {Petr-Gotzens}, M.~G., {McCaughrean}, M.~J., {Bouvier}, J.,
  {Duch{\^e}ne}, G., {Quirrenbach}, A., \& {Zinnecker}, H. 2006, \aap, 458, 461

\bibitem[{{Konopacky} {et~al.}(2007){Konopacky}, {Ghez}, {Rice}, \&
  {Duch{\^e}ne}}]{Konopacky:2007pd}
{Konopacky}, Q.~M., {Ghez}, A.~M., {Rice}, E.~L., \& {Duch{\^e}ne}, G. 2007,
  \apj, 663, 394

\bibitem[{{Kraus} \& {Hillenbrand}(2007{\natexlab{a}})}]{Kraus:2007ve}
{Kraus}, A.~L., \& {Hillenbrand}, L.~A. 2007{\natexlab{a}}, \apj, 662, 413

\bibitem[{{Kraus} \& {Hillenbrand}(2007{\natexlab{b}})}]{Kraus:2007mz}
---. 2007{\natexlab{b}}, \aj, 134, 2340

\bibitem[{{Kraus} \& {Hillenbrand}(2007{\natexlab{c}})}]{Kraus:2007gf}
---. 2007{\natexlab{c}}, \apj, 664, 1167

\bibitem[{{Kraus} \& {Hillenbrand}(2008)}]{Kraus:2008fr}
---. 2008, \apjl, 686, L111

\bibitem[{{Kraus} \& {Hillenbrand}(2009{\natexlab{a}})}]{Kraus:2009fk}
---. 2009{\natexlab{a}}, \apj, 704, 531

\bibitem[{{Kraus} \& {Hillenbrand}(2009{\natexlab{b}})}]{Kraus:2009uq}
---. 2009{\natexlab{b}}, \apj, 703, 1511

\bibitem[{{Kraus} {et~al.}(2011){Kraus}, {Ireland}, {Martinache}, \&
  {Hillenbrand}}]{Kraus:2011tg}
{Kraus}, A.~L., {Ireland}, M.~J., {Martinache}, F., \& {Hillenbrand}, L.~A.
  2011, \apj, 731, 8

\bibitem[{{Kraus} {et~al.}(2008){Kraus}, {Ireland}, {Martinache}, \&
  {Lloyd}}]{Kraus:2008zr}
{Kraus}, A.~L., {Ireland}, M.~J., {Martinache}, F., \& {Lloyd}, J.~P. 2008,
  \apj, 679, 762

\bibitem[{{Kraus} {et~al.}(2005){Kraus}, {White}, \&
  {Hillenbrand}}]{Kraus:2005qf}
{Kraus}, A.~L., {White}, R.~J., \& {Hillenbrand}, L.~A. 2005, \apj, 633, 452

\bibitem[{{Kraus} {et~al.}(2006){Kraus}, {White}, \&
  {Hillenbrand}}]{Kraus:2006ul}
---. 2006, \apj, 649, 306

\bibitem[{{Kroupa} {et~al.}(1999){Kroupa}, {Petr}, \&
  {McCaughrean}}]{Kroupa:1999fd}
{Kroupa}, P., {Petr}, M.~G., \& {McCaughrean}, M.~J. 1999, NewA, 4, 495

\bibitem[{{Lada} \& {Lada}(2003)}]{Lada:2003qo}
{Lada}, C.~J., \& {Lada}, E.~A. 2003, \araa, 41, 57

\bibitem[{{Lafreni{\`e}re} {et~al.}(2008){Lafreni{\`e}re}, {Jayawardhana},
  {Brandeker}, {Ahmic}, \& {van Kerkwijk}}]{Lafreniere:2008fq}
{Lafreni{\`e}re}, D., {Jayawardhana}, R., {Brandeker}, A., {Ahmic}, M., \& {van
  Kerkwijk}, M.~H. 2008, \apj, 683, 844

\bibitem[{{Law} {et~al.}(2010){Law}, {Dhital}, {Kraus}, {Stassun}, \&
  {West}}]{Law:2010vn}
{Law}, N.~M., {Dhital}, S., {Kraus}, A., {Stassun}, K.~G., \& {West}, A.~A.
  2010, \apj, 720, 1727

\bibitem[{{Leinert} {et~al.}(1993){Leinert}, {Zinnecker}, {Weitzel},
  {Christou}, {Ridgway}, {Jameson}, {Haas}, \& {Lenzen}}]{Leinert:1993gd}
{Leinert}, C., {Zinnecker}, H., {Weitzel}, N., {Christou}, J., {Ridgway},
  S.~T., {Jameson}, R., {Haas}, M., \& {Lenzen}, R. 1993, \aap, 278, 129

\bibitem[{{Lodieu} {et~al.}(2008){Lodieu}, {Hambly}, {Jameson}, \&
  {Hodgkin}}]{Lodieu:2008zg}
{Lodieu}, N., {Hambly}, N.~C., {Jameson}, R.~F., \& {Hodgkin}, S.~T. 2008,
  \mnras, 383, 1385

\bibitem[{{Luhman}(2004)}]{Luhman:2004yq}
{Luhman}, K.~L. 2004, \apj, 602, 816

\bibitem[{{Luhman}(2006)}]{Luhman:2006dp}
---. 2006, \apj, 645, 676

\bibitem[{{Luhman} {et~al.}(2003){Luhman}, {Stauffer}, {Muench}, {Rieke},
  {Lada}, {Bouvier}, \& {Lada}}]{Luhman:2003pb}
{Luhman}, K.~L., {Stauffer}, J.~R., {Muench}, A.~A., {Rieke}, G.~H., {Lada},
  E.~A., {Bouvier}, J., \& {Lada}, C.~J. 2003, \apj, 593, 1093

\bibitem[{{Mamon} \& {Soneira}(1982)}]{Mamon:1982vn}
{Mamon}, G.~A., \& {Soneira}, R.~M. 1982, \apj, 255, 181

\bibitem[{{Maxted} {et~al.}(2008){Maxted}, {Jeffries}, {Oliveira}, {Naylor}, \&
  {Jackson}}]{Maxted:2008gl}
{Maxted}, P.~F.~L., {Jeffries}, R.~D., {Oliveira}, J.~M., {Naylor}, T., \&
  {Jackson}, R.~J. 2008, \mnras, 385, 2210

\bibitem[{{Metchev} \& {Hillenbrand}(2009)}]{Metchev:2009hh}
{Metchev}, S.~A., \& {Hillenbrand}, L.~A. 2009, \apjs, 181, 62

\bibitem[{{Padgett} {et~al.}(1999){Padgett}, {Brandner}, {Stapelfeldt},
  {Strom}, {Terebey}, \& {Koerner}}]{Padgett:1999rt}
{Padgett}, D.~L., {Brandner}, W., {Stapelfeldt}, K.~R., {Strom}, S.~E.,
  {Terebey}, S., \& {Koerner}, D. 1999, \aj, 117, 1490

\bibitem[{{Patience} {et~al.}(2002){Patience}, {Ghez}, {Reid}, \&
  {Matthews}}]{Patience:2002qh}
{Patience}, J., {Ghez}, A.~M., {Reid}, I.~N., \& {Matthews}, K. 2002, \aj, 123,
  1570

\bibitem[{{Petr} {et~al.}(1998){Petr}, {Coud{\'e} du Foresto}, {Beckwith},
  {Richichi}, \& {McCaughrean}}]{Petr:1998ys}
{Petr}, M.~G., {Coud{\'e} du Foresto}, V., {Beckwith}, S.~V.~W., {Richichi},
  A., \& {McCaughrean}, M.~J. 1998, \apj, 500, 825

\bibitem[{{Preibisch} {et~al.}(2002){Preibisch}, {Brown}, {Bridges},
  {Guenther}, \& {Zinnecker}}]{Preibisch:2002qt}
{Preibisch}, T., {Brown}, A.~G.~A., {Bridges}, T., {Guenther}, E., \&
  {Zinnecker}, H. 2002, \aj, 124, 404

\bibitem[{{Radigan} {et~al.}(2009){Radigan}, {Lafreni{\`e}re}, {Jayawardhana},
  \& {Doyon}}]{Radigan:2009aj}
{Radigan}, J., {Lafreni{\`e}re}, D., {Jayawardhana}, R., \& {Doyon}, R. 2009,
  \apj, 698, 405

\bibitem[{{Reid} {et~al.}(2002){Reid}, {Gizis}, \& {Hawley}}]{Reid:2002sn}
{Reid}, I.~N., {Gizis}, J.~E., \& {Hawley}, S.~L. 2002, \aj, 124, 2721

\bibitem[{{Reid} {et~al.}(2003){Reid}, {Cruz}, {Allen}, {Mungall}, {Kilkenny},
  {Liebert}, {Hawley}, {Fraser}, {Covey}, \& {Lowrance}}]{Reid:2003fr}
{Reid}, I.~N., {et~al.} 2003, \aj, 126, 3007

\bibitem[{{Reipurth} \& {Clarke}(2001)}]{Reipurth:2001ec}
{Reipurth}, B., \& {Clarke}, C. 2001, \aj, 122, 432

\bibitem[{{Sartoretti} {et~al.}(1998){Sartoretti}, {Brown}, {Latham}, \&
  {Torres}}]{Sartoretti:1998bd}
{Sartoretti}, P., {Brown}, R.~A., {Latham}, D.~W., \& {Torres}, G. 1998, \aap,
  334, 592

\bibitem[{{Schlegel} {et~al.}(1998){Schlegel}, {Finkbeiner}, \&
  {Davis}}]{Schlegel:1998yj}
{Schlegel}, D.~J., {Finkbeiner}, D.~P., \& {Davis}, M. 1998, \apj, 500, 525

\bibitem[{{Schmidt-Kaler}(1982)}]{Schmidt-Kaler:1992ab}
{Schmidt-Kaler}, T. 1982, Landolt-Bornstein Numerical Data and Functional
  Relationships in Science and Technology, New Series, Group VI, Volume 2b,
  Springer-Verlag, Berlin

\bibitem[{{Siegel} {et~al.}(2002){Siegel}, {Majewski}, {Reid}, \&
  {Thompson}}]{Siegel:2002zr}
{Siegel}, M.~H., {Majewski}, S.~R., {Reid}, I.~N., \& {Thompson}, I.~B. 2002,
  \apj, 578, 151

\bibitem[{{Simon} {et~al.}(1995){Simon}, {Ghez}, {Leinert}, {Cassar}, {Chen},
  {Howell}, {Jameson}, {Matthews}, {Neugebauer}, \& {Richichi}}]{Simon:1995yi}
{Simon}, M., {et~al.} 1995, \apj, 443, 625

\bibitem[{{Slesnick} {et~al.}(2006{\natexlab{a}}){Slesnick}, {Carpenter}, \&
  {Hillenbrand}}]{Slesnick:2006pi}
{Slesnick}, C.~L., {Carpenter}, J.~M., \& {Hillenbrand}, L.~A.
  2006{\natexlab{a}}, \aj, 131, 3016

\bibitem[{{Slesnick} {et~al.}(2006{\natexlab{b}}){Slesnick}, {Carpenter},
  {Hillenbrand}, \& {Mamajek}}]{Slesnick:2006xr}
{Slesnick}, C.~L., {Carpenter}, J.~M., {Hillenbrand}, L.~A., \& {Mamajek},
  E.~E. 2006{\natexlab{b}}, \aj, 132, 2665

\bibitem[{{Slesnick} {et~al.}(2008){Slesnick}, {Hillenbrand}, \&
  {Carpenter}}]{Slesnick:2008mi}
{Slesnick}, C.~L., {Hillenbrand}, L.~A., \& {Carpenter}, J.~M. 2008, \apj, 688,
  377

\bibitem[{{Sterzik} {et~al.}(2003){Sterzik}, {Durisen}, \&
  {Zinnecker}}]{Sterzik:2003fx}
{Sterzik}, M.~F., {Durisen}, R.~H., \& {Zinnecker}, H. 2003, \aap, 411, 91

\bibitem[{{Stetson}(1987)}]{Stetson:1987ya}
{Stetson}, P.~B. 1987, \pasp, 99, 191

\bibitem[{{Szebehely} \& {Zare}(1977)}]{Szebehely:1977mz}
{Szebehely}, V., \& {Zare}, K. 1977, \aap, 58, 145

\bibitem[{{Tobin} {et~al.}(2009){Tobin}, {Hartmann}, {Furesz}, {Mateo}, \&
  {Megeath}}]{Tobin:2009fo}
{Tobin}, J.~J., {Hartmann}, L., {Furesz}, G., {Mateo}, M., \& {Megeath}, S.~T.
  2009, \apj, 697, 1103

\bibitem[{{Torres} {et~al.}(2009){Torres}, {Loinard}, {Mioduszewski}, \&
  {Rodr{\'{\i}}guez}}]{Torres:2009ct}
{Torres}, R.~M., {Loinard}, L., {Mioduszewski}, A.~J., \& {Rodr{\'{\i}}guez},
  L.~F. 2009, \apj, 698, 242

\bibitem[{{White} \& {Ghez}(2001)}]{White:2001jf}
{White}, R.~J., \& {Ghez}, A.~M. 2001, \apj, 556, 265

\bibitem[{{Wizinowich} {et~al.}(2006){Wizinowich}, {Le Mignant}, {Bouchez},
  {Campbell}, {Chin}, {Contos}, {van Dam}, {Hartman}, {Johansson}, {Lafon},
  {Lewis}, {Stomski}, {Summers}, {Brown}, {Danforth}, {Max}, \&
  {Pennington}}]{Wizinowich:2006zn}
{Wizinowich}, P.~L., {et~al.} 2006, \pasp, 118, 297

\end{thebibliography}

\clearpage

\LongTables

\begin{deluxetable}{lllllrrrl}
\tabletypesize{\scriptsize}
\tablewidth{0pt}
\tablecaption{Observed Sample}
\tablehead{\colhead{Name} & \colhead{RA} & \colhead{DEC} & 
\colhead{SpT} & \colhead{Mass} & \colhead{$K$} & 
\colhead{$R_{TT}$} & \colhead{$d_{TT}$} & \colhead{Refs}
\\
\colhead{} & \multicolumn{2}{c}{(J2000)} & \colhead{} & 
\colhead{($M_{\sun}$)} & \colhead{(mag)} & \colhead{(mag)} & 
\colhead{(arcsec)}
}
\startdata
\multicolumn{9}{c}{Taurus}\\
SCH J0359099+2009362 &3 59 09.85&+20 09 36.3&M4.75&0.20&12.53&15.2&47.4&\\
2MASS J04080782+2807280&4 08 07.82&+28 07 28.0&M3.75&0.30&11.39&12.0&9.1&\\
2MASS J04141188+2811535&4 14 11.88&+28 11 53.5&M6.25&0.09&11.64&10.0&23.3&\\
2MASS J04152409+2910434&4 15 24.09&+29 10 43.4&M7&0.058&12.36&13.4&45.8&\\
2MASS J04161885+2752155&4 16 18.85&+27 52 15.5&M6.25&0.09&11.35&16.1&28.1&\\
SCH J0416272+2053093&4 16 27.20&+20 53 09.3&M5&0.18&11.11&14.7&21.7&\\
2MASS J04163911+2858491&4 16 39.11&+28 58 49.1&M5.5&0.14&11.28&17.6&19.1&\\
\enddata
 \tablecomments{The properties of our sample members are summarized in our previous compilation, \citet[][]{Kraus:2007ve}. References for sources observed in previous multiplicity surveys:
 1) \citet[][]{Kraus:2006ul}
 ; 2) \citet[][]{Konopacky:2007pd}
 ; 3) \citet[][]{Kraus:2005qf}
 }
 \end{deluxetable}

\clearpage

\begin{deluxetable}{llllrr}
\tabletypesize{\scriptsize}
\tablewidth{0pt}
\tablecaption{Observations}
\tablehead{\colhead{Target} & \colhead{Epoch} &
\colhead{Mode} & \colhead{$t_{int}$} & \colhead{PSF FWHM} & \colhead{PSF Elongation}
\\
\colhead{} & \colhead{(JD-2450000)} & 
\colhead{} & \colhead{(s)} & \colhead{(mas)} & \colhead{(\%)}
}
\startdata
SCH J0359099+2009362 &3728&L&250&76&9\\
V410 X-ray 3&3728&L&150&75&31\\
2MASS J04213460+2701388&3728&L&150&56&26\\
2MASS J04141188+2811535&3728&N&150&87&11\\
MHO-Tau 8&3729&L&150&122&20\\
2MASS J04554757+3028077\tablenotemark{b}&3729&L&150&63&15\\
CFHT-Tau 16&3729&L&150&94&32\\
CFHT-Tau 13&3729&L&150&88&27\\
\enddata
\tablenotetext{a}{Observation also includes 2M04554801+3028050 in the same field.}
\end{deluxetable}

\clearpage

\begin{deluxetable}{lcrccc}
\tabletypesize{\scriptsize}
\tablewidth{0pt}
\tablecaption{Candidate Companions to Young Stars and Brown Dwarfs}
\tablehead{\colhead{Target} & \colhead{Epoch} &
\colhead{$N_{meas}$} & \colhead{Sep} & \colhead{PA} & 
\colhead{$\Delta$$K'$}
\\
\colhead{} & \colhead{(JD-2450000)} & 
\colhead{} & \colhead{(mas)} & \colhead{(deg)} & \colhead{(mag)}
}
\startdata
\multicolumn{6}{c}{Taurus}\\
2MASS J04080782+2807280 B\tablenotemark{a}&4070&2&52$\pm$1&6.74$\pm$1.63&0.03$\pm$0.03\\
2MASS J04152409+2910434 c1&4067&3&4521$\pm$5&0.29$\pm$0.01&5.72$\pm$0.02\\
2MASS J04163911+2858491 B&4068&3&214$\pm$1&218.6$\pm$0.5&1.14$\pm$0.03\\
2MASS J04163911+2858491 B&4089&5&210$\pm$1&219.4$\pm$0.2&1.06$\pm$0.01\\
2MASS J04202555+2700355 c1&3728&2&4621$\pm$5&267.498$\pm$0.004&6.94$\pm$0.12\\
2MASS J04230607+2801194 c1&4068&1&6443$\pm$6&291.63$\pm$0.02&4.79$\pm$0.05\\
2MASS J04284263+2714039 B\tablenotemark{a}&3728&6&627$\pm$1&350.63$\pm$0.04&0.461$\pm$0.004\\
\enddata
\tablecomments{We label a companion as ``B'' if falls in the range of parameter space where background stars are statistically unlikely to have been found ($K < 14$, $\rho < 3$\arcsec; Section 4.2), and otherwise label them as ``c\#'' to reflect that they are unconfirmed candidate companions. For wide pairs, we refer to a specific component as ``A'' or ``B'', and then refer to the newly-identified (closer) companion to that component as ``b'' or ``c1'' following the same rules.}
\tablenotetext{a}{Previously identified as a candidate companion by Ghez et al. (1993) (V928 Tau), Luhman (2004) (2MASS J04554757+3028077), Kraus et al. (2005) (USco 55, USco 66), Kraus et al. (2006) (V410 X-ray 3), Konopacky et al. (2007) (2MASS J04284263+2714039 and 2MASS J04403979+2519061), Kraus \& Hillenbrand (2007) (USco J160702.1-201938, USco J160904.0-193359, and USco J160908.4-200928), and Kraus et al. (2011) (2MASS J04080782+2807280).}
\end{deluxetable}

\clearpage

\begin{landscape}
\begin{deluxetable}{lccccccccccccccc}
\tabletypesize{\scriptsize}
\tablewidth{0pt}
\tablecaption{Companion Detection Limits}
\tablehead{
\colhead{Target} & \multicolumn{15}{c}{$\Delta$$K'$ (mag) at $\rho$$=$ 
(mas)}
\\
\colhead{} &
\colhead{40} & \colhead{45} & \colhead{50} & 
\colhead{60} & \colhead{80} & \colhead{100} & 
\colhead{120} & \colhead{150} & \colhead{200} & 
\colhead{300} & \colhead{400} & \colhead{500} & 
\colhead{750} & \colhead{1000} & \colhead{2000}
}
\startdata
\multicolumn{16}{c}{Taurus}\\
2MASS J04080782+2807280&...&...&...&...&...&...&...&0.84&1.32&2.07&2.79&3.51&5.01&5.88&7.56\\
2MASS J04141188+2811535&...&...&...&0.55&0.93&0.97&1.01&1.07&1.17&1.37&1.57&2.21&3.52&4.31&6.25\\
2MASS J04141188+2811535&...&0.67&0.81&1.09&1.19&1.29&1.39&2.24&3.07&3.73&4.23&4.88&6.34&6.76&7.74\\
2MASS J04141188+2811535&0.66&0.81&0.97&1.28&1.50&1.73&1.96&2.51&3.71&3.97&5.12&5.69&6.39&6.92&7.28\\
2MASS J04152409+2910434&0.71&0.87&1.03&1.35&1.60&1.86&2.12&2.54&2.45&3.52&4.60&5.39&6.29&7.10&8.06\\
2MASS J04161885+2752155&...&...&...&0.68&1.10&1.37&1.63&2.09&2.60&3.05&3.78&4.34&5.46&6.25&7.94\\
2MASS J04163911+2858491&...&...&...&...&...&...&...&...&0.78&1.30&2.17&2.77&4.06&5.03&7.70\\
\enddata
\end{deluxetable}

\clearpage
\end{landscape}

\begin{deluxetable}{llrrll}
\tabletypesize{\scriptsize}
\tablewidth{0pt}
\tablecaption{Binary System Properties}
\tablehead{
\colhead{Target Name} & \colhead{Epoch} & \colhead{Proj. Sep.} & \colhead{$q$} & 
\colhead{$M_{prim}$} & \colhead{$M_{sec}$}
\\
\colhead{} & \colhead{(JD-2450000)} & \colhead{(AU)} & \colhead{($M_s/M_p$)} & 
\colhead{($M_{\odot}$)} & \colhead{($M_{\odot}$)}
}
\startdata
\multicolumn{6}{c}{Taurus}\\
2MASS J04080782+2807280 A-B&4070&7.5&0.98&0.30&0.29\\
2MASS J04163911+2858491 A-B&4089&31&0.40&0.14&0.056\\
2MASS J04163911+2858491 A-B&4068&31&0.38&0.14&0.054\\
2MASS J04284263+2714039 A-B&3728&91&0.60&0.16&0.095\\
2MASS J04284263+2714039 A-B&3772&91&0.48&0.16&0.076\\
2MASS J04403979+2519061 A-B&4070&7.1&1.00&0.16&0.16\\
2MASS J04554757+3028077 A-B&3728&920&0.22&0.20&0.044\\
2MASS J04554757+3028077 A-B&3773&930&0.21&0.20&0.041\\
2MASS J04554970+3019400 A-B&4067&8.1&0.95&0.10&0.095\\
V410 X-ray 3 A-B&3773&7.0&0.77&0.08&0.062\\
V410 X-ray 3 A-B&3728&7.1&0.80&0.08&0.064\\
V928 Tau A-B&4068&32&0.97&0.60&0.58\\
\multicolumn{6}{c}{Upper Sco}\\
SCH J15582566-18260865 A-B&4186&14&0.98&0.074&0.072\\
USco 55 A-B&3772&18&0.97&0.10&0.097\\
USco 66 A-B&3772&12&0.89&0.074&0.066\\
USco 80 Aa-Ab&4187&7.8&0.97&0.36&0.35\\
USco J160702.1-201938 A-B&3772&220&0.50&0.13&0.064\\
USco J160702.1-201938 Aa-Ab&3772&8.0&0.91&0.13&0.12\\
USco J160904.0-193359 A-B&3772&190&0.84&0.24&0.20\\
USco J160908.4-200928 A-B&3773&300&0.87&0.24&0.21\\
\enddata
\end{deluxetable}

\clearpage

\begin{deluxetable}{lrrcllccl}
\tabletypesize{\tiny}
\tablewidth{0pt}
\tablecaption{Binary Companions from Previous Surveys}
\tablehead{
\colhead{Target} & 
\colhead{RA} & 
\colhead{DEC} & 
\colhead{Sep} & 
\colhead{$\Delta$$m$} &
\colhead{$M_{prim}$} & 
\colhead{Sep} & 
\colhead{$q$} &
\colhead{Refs}
\\
\colhead{} & \multicolumn{2}{c}{(J2000)} & 
\colhead{(mas)} & 
\colhead{(mag)} & 
\colhead{($M_{\sun}$)} & 
\colhead{(AU)} & 
\colhead{($M_s/M_p$)}
}
\startdata
\multicolumn{9}{c}{Cha-I}\\
T5&10 57 42.20&-76 59 35.7&159$\pm$1&$\Delta K=0.05 \pm 0.01$&0.36&27&0.97&1\\
CHXR 71&11 02 32.65&-77 29 13.0&572$\pm$1&$\Delta K=1.63 \pm 0.01$&0.40&97&0.23&1\\
Hn 4&11 05 14.67&-77 11 29.1&211$\pm$1&$\Delta K=0.04 \pm 0.01$&0.36&36&0.97&1\\
CHXR 15&11 05 43.00&-77 26 51.7&304$\pm$1&$\Delta K=0.65 \pm 0.01$&0.16&52&0.53&1, 2\\
CHXR 26&11 07 36.87&-77 33 33.5&1396$\pm$1&$\Delta K=0.47 \pm 0.01$&0.33&240&0.68&1\\
ChaH$\alpha$ 2&11 07 42.45&-77 33 59.4&167$\pm$1&$\Delta K=0.11 \pm 0.02$&0.16&28&0.90&1, 2\\
T43&11 09 54.08&-76 29 25.3&796$\pm$1&$\Delta K=1.41 \pm 0.01$&0.50&135&0.30&1\\
\enddata
\tablecomments{References: 
1) \citet[][]{Lafreniere:2008fq};
2) \citet[][]{Ahmic:2007hn};
3) \citet[][]{Kraus:2007gf};
4) \citet[][]{Briceno:2002jo};
5) \citet[][]{Ghez:1993xh};
6) \citet[][]{White:2001jf} ;
7) \citet[][]{Simon:1995yi};
8) \citet[][]{Konopacky:2007pd};
9) \citet[][]{Kraus:2007ve};
10) \citet[][]{Kraus:2009uq};
11) \citet[][]{Kraus:2006ul};
12) \citet[][]{Kohler:2000lo};
13) \citet[][]{Kraus:2005qf};
14) \citet[][]{Kraus:2008zr};
15) \citet[][]{Biller:2011mw}.
}
\end{deluxetable}

\clearpage

\begin{landscape}

\begin{deluxetable}{lrrlcccccccccccccccccl}
\tabletypesize{\tiny}
\tablewidth{0pt}
\tablecaption{Companion Detection Limits from Previous Surveys}
\tablehead{
\colhead{Target} & \colhead{RA} & \colhead{DEC} & \colhead{$M_{prim}$} &
\multicolumn{17}{c}{$q=M_s/M_p$ at $\log{\rho}$$=$ (AU)} & 
\colhead{Refs}
\\
\colhead{} & \multicolumn{2}{c}{(J2000)} & \colhead{($M_{\sun}$)} & 
\colhead{0.5} & \colhead{0.6} & \colhead{0.7} & 
\colhead{0.8} & \colhead{0.9} & \colhead{1.0} & 
\colhead{1.1} & \colhead{1.2} & \colhead{1.3} & 
\colhead{1.4} & \colhead{1.6} & \colhead{1.8} & 
\colhead{2.1} & \colhead{2.4} & \colhead{2.7} & 
\colhead{3.0} & \colhead{3.6}  
}
\startdata
\multicolumn{22}{c}{Cha-I}\\
T5 A&10 57 42.20&-76 59 35.7&0.36&...&...&...&...&...&...&...&0.29&0.08&0.07&0.04&0.04&...&...&...&...&...&1\\
T10&11 00 40.22&-76 19 28.1&0.30&...&...&...&...&...&...&...&0.33&0.12&0.11&0.08&0.07&0.07&0.07&0.07&...&...&1\\
CHXR9C A&11 01 18.75&-76 27 02.5&0.48&...&...&...&...&...&...&...&0.34&0.14&0.11&0.05&0.04&0.04&0.04&0.04&...&...&1\\
2MASS1110-7722 A&11 01 19.26&-77 32 38.3&0.27&...&...&...&...&...&...&...&0.33&0.13&0.11&...&...&...&...&...&...&...&1\\
2MASS1110-7722 A+B&11 01 19.26&-77 32 38.3&0.41&...&...&...&...&...&...&...&...&...&...&0.05&0.03&0.03&0.03&0.03&...&...&1\\
CHXR71 A&11 02 32.65&-77 29 13.0&0.40&...&...&...&...&...&...&...&0.30&0.08&0.07&0.05&0.04&0.04&...&...&...&...&1\\
CHXR71 A+B&11 02 32.65&-77 29 13.0&0.49&...&...&...&...&...&...&...&...&...&...&...&...&...&0.03&0.03&...&...&1\\
\enddata
\tablecomments{The properties of most targets, including the spectral types used to estimate their masses, are summarized in our previous compilation, \citet[][]{Kraus:2007ve}. Properties for the Upper Sco members discovered by \citet[][]{Lodieu:2008zg} are listed in that work. References for past high-resolution imaging observations:
1) \citet[][]{Lafreniere:2008fq};
2) this work;
3) \citet[][]{Kraus:2007ve};
4) \citet[][]{Kraus:2009uq};
5) \citet[][]{Ghez:1993xh};
6) \citet[][]{Biller:2011mw};
7) \citet[][]{Kraus:2006ul};
8) \citet[][]{Konopacky:2007pd};
9) \citet[][]{Sartoretti:1998bd};
10) \citet[][]{Simon:1995yi};
11) \citet[][]{Duchene:2007uq};
12) \citet[][]{Kohler:2000lo};
13) \citet[][]{Kraus:2005qf}.
}
\end{deluxetable}

\clearpage

\end{landscape}

\end{document}